\documentclass[useAMS,usenatbib,aas_macros]{mn2e} 
\usepackage {graphicx,times,aas_macros}

\newcommand{\uno}{{\rm B}}
\newcommand{\dos}{{\rm C}}
\newcommand{\tres}{{\rm A}}

\newcommand{\equ}[1]{eq.~(\ref{eq:#1})}

\newcommand{\se}[1]{\S\ref{sec:#1}}
\newcommand{\fig}[1]{Fig.~\ref{fig:#1}}

\newcommand{\Fig}[1]{Figure~\ref{fig:#1}}

\newcommand{\be}{\begin{equation}}
\newcommand{\ee}{\end{equation}}
\newcommand{\bea}{\begin{eqnarray}}
\newcommand{\eea}{\end{eqnarray}}

\newcommand{\msun}{{\rm M}_\odot}
\newcommand{\Msun}{M_\odot}

\newcommand{\lsun}{{\rm L}_\odot}

\newcommand{\ifm}[1]{\relax\ifmmode#1\else$\mathsurround=0pt #1$\fi}
\newcommand{\kms}{\ifmmode\,{\rm km}\,{\rm s}^{-1}\else km$\,$s$^{-1}$\fi}
\newcommand{\hmpc}{\,\ifm{h^{-1}}{\rm Mpc}}

\newcommand{\kpc}{\,{\rm kpc}}
\newcommand{\pc}{\,{\rm pc}}
\newcommand{\Gyr}{\,{\rm Gyr}}

\newcommand{\Myr}{\,{\rm Myr}}

\newcommand{\ltsima}{$\; \buildrel < \over \sim \;$}
\newcommand{\lsim}{\lower.5ex\hbox{\ltsima}}
\newcommand{\gtsima}{$\; \buildrel > \over \sim \;$}
\newcommand{\gsim}{\lower.5ex\hbox{\gtsima}}

\def\omm{\Omega_{\rm m}}
\def\oml{\Omega_{\Lambda}}
\def\omb{\Omega_{\rm b}}
\def\sy{\,M_\odot\, {\rm yr}^{-1}}

\def\cmc{\,{\rm cm}^{-3}}

\def\Mv{M_{\rm v}}
\def\Rv{R_{\rm v}}
\def\Vv{V_{\rm v}}

\def\Md{M_{\rm d}}
\def\Mc{M_{\rm c}}
\def\Rc{R_{\rm c}}
\def\Rd{R_{\rm d}}
\def\Rb{R_{\rm b}}
\def\Mt{M_{\rm tot}}
\def\Mbd{\dot{M}_{\rm b}}
\def\fb{f_{\rm b}}
\def\Ms{M_*}

\def\td{t_{\rm d}}
\def\tm{t_{\rm mig}}
\def\te{t_{\rm enc}}
\def\tevac{t_{\rm evac}}
\def\tdis{t_{\rm dis}}
\def\ta{t_{\rm acc}}

\def\Qc{Q_{\rm c}}
\def\Qs{Q_{\rm s}}
\def\Qg{Q_{\rm g}}

\def\sigs{\sigma_{\rm s}}
\def\sigg{\sigma_{\rm g}}
\def\Sigs{\Sigma_{\rm s}}
\def\Sigg{\Sigma_{\rm g}}
\def\siggs{\sigma_{\rm gs}}

\def\lamj{\lambda_{\rm J}}
\def\cs{c_{\rm s}}

\def\Pf{P_{\rm floor}}
\def\Nc{N_{\rm c}}
\def\fj{f_J}

\usepackage{color}

\begin{document}

\large

\title[High-redshift Clumpy Discs in Simulations]{High-Redshift Clumpy Discs 
\& Bulges in Cosmological Simulations}

\author[D. Ceverino, A. Dekel \& F. Bournaud]
{Daniel Ceverino$^1$\thanks{E-mail: ceverino@phys.huji.ac.il}, 
Avishai Dekel$^1$\thanks{E-mail: dekel@phys,huji.ac.il} 
and Frederic Bournaud$^2$\thanks{E-mail: frederic.bournaud@cea.fr}\\ 
$^1$Racah Institute of Physics, The Hebrew University, Jerusalem 91904,
Israel\\
$^2$CEA, IRFU, SAp, 91191 Gif-sur-Yvette, France}

\date{}

\pagerange{\pageref{firstpage}--\pageref{lastpage}} \pubyear{0000}

\maketitle

\label{firstpage}

\begin{abstract}
We analyze the first
cosmological simulations that recover the fragmentation of 
high-redshift galactic discs driven by cold streams.
The fragmentation is recovered owing to an AMR resolution 
better than $70\pc$ with cooling below $10^4$K. 
We study three typical star-forming galaxies in haloes of 
$\sim 5\times 10^{11}\msun$ 
at $z \simeq 2.3$, when they were not undergoing a major merger.
The steady gas supply by cold streams leads to gravitationally unstable, 
turbulent discs, which fragment into giant clumps and transient features on a 
dynamical timescale. The disc clumps are not associated with dark-matter haloes.
The clumpy discs are self-regulated by gravity in a marginaly unstable 
state. 
Clump migration and angular-momentum transfer on an orbital timescale help the 
growth of a central bulge with a mass comparable to the disc.  The continuous 
gas input keeps the system of clumpy disc and bulge in a near {\it steady 
state\,} for several Gyr. 
The average star-formation rate, much of which occurs in the
clumps, follows the gas accretion rate of $\sim 45 \sy$.
The simulated galaxies resemble in many ways the observed star-forming 
galaxies at high redshift.
Their properties are consistent with the simple theoretical framework 
presented in \citet[][DSC]{dsc}.  
In particular, a two-component analysis reveals that the simulated 
discs are indeed marginally unstable, and the time evolution confirms the 
robustness of the clumpy configuration in a cosmological steady state.  
By $z \sim 1$, the simulated systems are stabilized by a dominant stellar 
spheroid, demonstrating the process of ``morphological quenching" of star 
formation \citep{Martig09}.   
We demonstrate that the disc fragmentation 
is not a numerical artifact once 
the Jeans length is kept larger than $\sim7$ resolution elements, i.e. beyond 
the standard Truelove criterion.
\end{abstract}


\section{Introduction}
\label{sec:intro}

Observations at high redshifts in the range $z \sim 1-3$ 
have opened a new window into the most active period of galaxy formation 
\citep{Steidel99,Adelberger04,Daddi04,
Genzel06,Forster06,Elmegreen06,Genzel08,Stark08,Law09,Forster09}. 
Both observation and theory indicate that this was a rather violent process. 
Continuous, intense, cold gas streams along the cosmic web
fueled dense gaseous discs and induced high star formation rates (SFR)
on the order of $\sim100 \sy$ 
\citep{dekel09},
much higher than the few $\sy$ in today's galaxies.
We refer to these galaxies as SFGs, standing for ``Star-Forming Galaxies" 
and for ``Stream-Fed Galaxies".
In these extreme conditions, the images of most SFGs are irregular
with giant clumps \citep{Cowie95,Bergh96,Elmegreen05,Elmegreen07}. 
Although some of these irregular morphologies may be 
associated with mergers or galaxy 
interactions, the morphology and kinematics of many of the SFGs indicates
that they are rotationally supported, extended discs 
\citep{Genzel06,Forster06,Genzel08}.
It seems that most SFGs are incompatible with being ongoing major mergers 
or remnants of such mergers 
\citep{Shapiro08,Bournaud08,dekel09,Bournaud09,dsc}, 
though there are counter examples \citep[e.g.,][]{RobertsonBullock}.
This indicates a paradigm shift in our understanding of galaxy formation.

The high-redshift galactic discs are very different from the 
low-redshift disc galaxies.
In a typical $z \sim 2$ disc, $10-40\%$ of its UV-restframe light 
is emitted from a few clumps of a characteristic size $\sim 1\kpc$, 
and with a high SFR of tens of $\sy$ each
\citep{Elmegreen04, ElmegreenE05, Forster06,Genzel08}.
These clumps represent giant star-forming regions, much larger than the 
star-forming complexes in local galaxies.
This pronounced clumpy morphology at high redshift is not
limited to the UV as the main high-redshift clumps are also seen in optical
rest-frame emission \citep{Genzel08,Forster09}. It is not a bandshift 
effect in the sense that the UV images of clumpy low-redshift galaxies 
would not appear so clumpy if put at high redshift and observed with 
limited resolution and signal-to-noise ratio \citep{Elmegreen09b}.
The clumpy galaxies where first dubbed ``clump-clusters" galaxies 
and ``chain" galaxies when viewed face on or edge on respectively 
\citep{Cowie95,Elmegreen04b}.
Most of the massive ones
are thick rotating discs, with high velocity dispersions of 
$\sigma = 20 - 80 \kms$ 
(one dimensional) and rotation to dispersion ratio of
$V/\sigma \sim 1-7$ \citep{Cresci09}, 
as opposed to today's thin discs of $\sigma \simeq 10\kms$.
The smaller galaxies contain a large fraction of dispersion-dominated systems
\citep{Law07,Law09,Forster09}.
Estimates of the total gas fraction in SFGs, based on CO measurements,
range from 0.2 to 0.8, with an average of $\sim 0.4-0.6$ 
\citep{Tacconi08,Daddi08,Daddi09},  
systematically higher than the typical gas fraction of $\sim 0.1$ in today's 
discs.
This gas consists of large molecular cloud complexes 
that host the observed star-forming regions.
The typical age of the stellar populations in these clumps 
are crudely estimated to range from 
one to several hundreds Myr
\citep{Elmegreen09,Forster09},
on the order of ten dynamical times, indicating that the clumps 
may plausibly
survive for such durations.

A substantial fraction
of these high-$z$ clumpy discs show a central stellar bulge
\citep{Genzel08,Elmegreen08c}. 
Other observations indicate that a large population of compact, massive
spheroids with suppressed SFR
is already present at those redshifts \citep{Kriek06,Dokkum08,Kriek09}.
This means that a well-developed Red Sequence was already in place at 
$z \sim$ 2. 
While major mergers do lead to spheroids
\citep{Robertson06,Cox06,DekelCox,Hopkins06}, 
the major-merger rate is not enough for producing spheroids
in the observed abundance 
\citep[][and references therein]{dekel09,Genel08,Stewart09},
so an additional and complementary mechanism 
must have been working efficiently at high redshifts.

The gravitational fragmentation of gas-rich and turbulent galactic discs 
into giant clumps, and their subsequent migration into a central bulge, 
have been addressed before in the simplified context of isolated galaxies 
\citep{Noguchi99, Elmegreen05,  Bournaud07, Genzel08, Elmegreen08a}. 
According to the standard Toomre instability analysis \citep{toomre64},
the material in a rotating disc becomes unstable to local gravitational 
collapse if its surface density is sufficiently high for its self-gravity to
overcome the forces induced by differential rotation and velocity 
dispersion that work against the collapse. 
If the disc is maintained in a marginally unstable state
with relatively high velocity dispersion,
the clumps are big. Then, the timescale for their migration to the centre 
by gravitational interactions is short, on the order of ten 
disc dynamical times, or a couple of orbital periods at the disc radius.  

The clumps and the other transient perturbations in the disc drive
further inflow of disc mass toward the centre through angular-momentum
transport outward.  Together they lead to a rapid growth of a bulge in 
about $0.5 \Gyr$, at the expense of the draining and stabilization of the disc.

The large fraction of clumpy massive galaxies at $z \sim 2$ 
\citep[e.g.,][]{Elmegreen07,  Tacconi08} 
suggests that the clumpy phase should last for a few Gigayears.
DSC have argued that when a galaxy is part of the high-redshift 
cosmological environment, it can naturally be in a long-term steady state 
typically consisting of a clumpy disc and a comparable bulge.
Massive galaxies at that epoch are fed by continuous, intense cold gas streams
that follow the filaments of the cosmic web 
\citep{keres05, db06, ocvirk08, dekel09}.
The fresh supply of cold gas refills the disc as it is being drained into the 
bulge, and keeps the surface density sufficiently high for disc instability 
and new clump formation as long as the bulge is not too massive. 
On the other hand, if a large fraction of the incoming streams is already in 
clumps, they can stabilize the disc against fragmentation 
in two complementary ways:
by rapid direct buildup of the bulge via mergers
and by generating high velocity dispersion in the disc. 
This led DSC to suggest that a low level of clumpiness in the streams is a 
key factor in maintaining a clumpy disc in steady state. 

This long-term clumpy phase
can be naturally addressed by cosmological, hydrodynamical simulations, 
though the numerical challenges are not trivial.
The dynamical range must explore the cosmological streams
on scales of hundreds of kpc while properly resolving the gas physics 
in the disc on scales of tens of parsecs.
These difficulties limited the earlier simulations of disc fragmentation 
to idealized discs in isolation 
\citep{Noguchi99, Immeli04a, Immeli04b, Bournaud07, Elmegreen08a, Bournaud09, Bournaud09b}.
They successfully reproduced the Toomre instability, 
the subsequent migration of clumps and the formation of a classical bulge.
However, in the absence of fresh cosmological gas supply, these phenomena 
were limited to one episode of instability and disc evacuation, and 
were not capable of capturing the long-term cosmological steady state. 
On the other hand, the earlier simulations that did follow the 
cosmological gas 
supply did not have the necessary resolution with proper gas physics
to reproduce the gravitational fragmentation of the discs 
\citep{Springel03, keres05, Governato07, ocvirk08, dekel09}. 

A key ingredient for recovering the fragmentation is allowing the gas to cool
to temperatures below $10^4$ K and reach densities well above $1$ atom
per cc, typical of gas in giant molecular clouds. 
If the gas is kept at higher temperatures, the pressure prevents 
local gravitational collapse. 
This may be fine in the modeling of present-day disks that do not
suffer from wild gravitational instability \citep{Springel03,Robertson04}, 
but it is not adequate when simulating the high-redshift gaseous and
continuously fed disks that are likely to develop wild disk instabilities.

Even with state-of-the-art resolution of $\sim 30$ pc,
where the gas can cool to $\sim 100$ K and reach $n \sim 100 \cmc$
\citep{Ceverino09, Agertz09b}, 
the thermal Jeans scale in the dense gas is below the resolution scale.
In other words, there is no pressure to support the gas against gravitational
collapse on the resolution scale, so the gas tends to fragment artificially
\citep{Truelove, Bate97}.
In order to prevent artificial fragmentation, the common procedure is
to introduce a pressure floor \citep{Machacek01,RobertsonKravtsov,Agertz09a}, 
but it has to be applied in a way that would not prevent real 
fragmentation on the turbulence Jeans scale when it is only a few
times the resolution scale.

In this paper we present the results from three different
high-resolution cosmological simulations, zooming in on three nearly 
random galaxies with haloes of 
$\sim (1-2)\times 10^{12}\msun$ at $z=1$.
They have reached $\sim 5\times 10^{11}\msun$ at $z=2.3$, 
where we perform a detailed analysis. 
In these simulations, we do allow cooling to temperatures well below $10^4$K
and thus allow for real physical disc fragmentation, while applying the 
pressure floor necessary for preventing artificial fragmentation. 
We find that they all show clumpy discs similar to the observations
and according to theory.
In \se{theory} we summarize the theoretical framework laid out in DSC.
In \se{models} we describe the simulation method. 
In \se{global} we address the global properties of the simulated galaxies
and verify that they obey the observed scaling relations at high redshift. 
In \se{results} we study in detail the three simulated galaxies at 
$z=2.3$ where we focus on the disc fragmentation and bulge formation, in
comparison with theory and observations.
\se{images} brings projected images of the different galaxy components,
\se{UDF} compares simulated and observed images,
\se{GalProp} describes the properties of the clumpy discs and bulges in
comparison to the theory as laid out in DSC,
\se{instability} refers to a two-component instability analysis,
\se{evo} follows the formation of disc clumps and their migration 
to the bulge,
and \se{steady} highlights the cosmological steady state of clumpy discs,
and the possible eventual stabilization by a massive spheroid.
In \se{artif} we test the robustness of our results to artificial 
fragmentation. 
Finally, in \se{disc} we summarize our results and discuss them.

\section{Theoretical Framework}
\label{sec:theory}

In a companion paper, DSC have laid out a simple theoretical framework for
clumpy discs driven by cosmological cold streams at high-redshift.
We summarize the theory here to help us interpret the simulation 
results as we present them. Note that the theoretical expectations did 
not affect the numerical experiments themselves in any way. 

Well before $z \sim 1$, the high accretion rate is expected to keep the 
discs gas rich, which validates a simplified one-component analysis as a first
crude approximation. 
According to the standard Toomre instability
analysis \citep[][Chapter 6]{toomre64,bt08}
a rotating disc becomes unstable to axi-symmetric modes once the local 
gravity overcomes both differential rotation and pressure
due to turbulence or thermal motions.
This is expressed in terms of the local stability parameter $Q$ being
smaller than a critical value of order unity,
\be
Q=\frac{\sigma_r \kappa}{\pi G \Sigma} < \Qc\, .
\label{eq:Q} 
\ee
Here $\Sigma$ is the surface density of the disc at radius $r$,  
$\sigma_r$ is the radial velocity dispersion,
(or the gas sound speed if it is larger),  
and $\kappa$ is the epicyclic frequency.
We adopt $\kappa=\sqrt{3}\Omega$, where $\Omega$ is the angular circular 
velocity.
The factor $\sqrt{3}$ stands for a number between 1 and 2, depending on
the shape of $\Omega(r)$.
For a thick disc, $\Qc \simeq 0.68$ \citep{goldreich65_thick}.

The disc of mass $\Md$ is expected to develop transient elongated sheared 
features and to fragment into a few bound massive clumps of a typical mass
\be
\Mc \simeq 0.27 \delta^2 \Md 
\label{eq:Mc}
\ee
and a typical radius 
\be
\Rc \sim 0.52 \delta \Rd \, ,
\label{eq:Rc}
\ee
where $\delta$ is the disc fraction of the total mass within the disc radius
$\Rd$,
\be
\delta \equiv \frac{\Md}{\Mt(\Rd)} \, . 
\label{eq:delta}
\ee
The fraction of disc mass in the clumps is crudely estimated to be
in the range $\alpha \sim 0.1-0.4$.
Note that a higher $\delta$ corresponds to more massive clumps.  

An unstable disc is expected to self-regulate itself at $Q \simeq \Qc$
with a velocity dispersion-to-rotation ratio
\be
\frac{\sigma_r}{V} \simeq 3^{-1/2} \Qc \delta \, .
\label{eq:sigma}
\ee
The turbulence can be largely maintained by the gravitational encounters of 
the clumps themselves, which operate on a timescale 
$\te \simeq 2.1 Q^2 \alpha^{-1} \td$,
where $\td \equiv \Omega^{-1} \simeq 50\Myr$ is the disc dynamical time.
With $Q \simeq 0.67$ and $\alpha \simeq 0.2$, this matches the timescale
for turbulence decay, $\tdis \simeq 1.4 Q^{-1} \td$.
The estimated effect of clump encounters is a lower limit to the total
effect of the gravitational interactions involving all the components 
of the perturbed disc, which maintain the marginally unstable configuration.
A higher $\delta$ thus corresponds to a higher $\sigma_r/V$ and a thicker disc.

The same gravitational encounters and dynamical friction make the giant clumps
migrate to the centre on a timescale 
\be
\tm \simeq 2.1 Q^2 \delta^{-2} \td 
\label{eq:tm}
\ee
and grow a bulge.
The associated evacuation timescale for the entire disc mass is
$\tevac \simeq 10.5 \alpha_{.2}^{-1} Q^2 \delta^{-2} \td$.
Additional comparable contributions to mass inflow in the disc
are associated with angular-momentum transport outward, partly induced
by the clump migration and partly due to torques involving the transient 
features.
We noted that a higher $\delta$ corresponds to more massive clumps. Then, 
With $Q \sim \Qc$, a higher $\delta$ or $\sigma_r/V$, namely more
massive clumps, correspond to a more rapid migration.

At high redshift, the continuous gas supply can maintain the disc unstable
with giant clumps for several Gigayears. 
The average cosmological gas accretion rate \citep{Neistein06,Genel08} 
is expected to be
\be
\dot{M} = 6\, M_{12}^{1.15}\, (1+z)^{2.25}\, f_{0.16} \sy \, 
\label{eq:Mdot}
\ee
where $M_{12}$ is the halo virial mass in units of $10^{12}\msun$,
and $f_{0.16}$ is the baryon fraction in units of $0.16$.
Based on the the effective deep
penetration seen in cosmological simulations \citep{dekel09},
the cosmological streams feed baryons into massive galaxies on a timescale
$\ta \simeq 44 \tau^{-1} \td$, with $\tau \simeq 1$ at $z = 2$ and varying
from 2.5 to 0.4 between $z\simeq 9$ and $1$.
The smooth component of the incoming streams, including small clumps,
replenishes the evacuating disc, while the massive clumps associated
with the incoming streams merge to the bulge.
If the galaxy  streams in which the massive clump fraction is less
than average,
the system is predicted to settle into a near steady state
with $\delta \simeq 0.3$, i.e., the input by streams and the transport from
disc to bulge maintain a constant bulge-to-disc ratio near unity.
This corresponds to comparable contributions of disc, bulge and 
dark matter to $\Mt(\Rd)$.

On the other hand,
in galaxies where the incoming streams are more clumpy than average,
the merging of external massive clumps 
(possibly associated with dark matter haloes)
 are expected to grow a dominant spheroid with
$\delta < 0.25$, and the dense clumps are capable of stirring up the 
turbulence in the disc to levels that stabilize the disc.  
DSC thus hypothesized that the dependence of the instability on the degree 
of clumpiness in the streams introduces a robust 
bimodality in the galaxy properties starting at $z \geq 3$.

If the properties of the streams feeding a given galaxy vary in time,
and in particular if the degree of clumpiness in these streams evolves, 
the galaxy may go through transitions from an unstable disc-dominated 
configuration to a stable bulge-dominated state and vice versa
\citep{Martig09}. 
The instability can be responsible for efficient star formation 
while the stability may result in a significant suppression of star formation.
However, after $z \sim 2$, the recovery from a bulge-dominated system
back to an unstable disc takes several Gigayear and may never materialize.

A systematic change in the stream properties is expected after $z \sim 1-2$.
Then, the cosmological accretion rate becomes slower,  
the smooth cold streams no longer penetrate very effectively through
the shock-heated media in massive haloes of $\sim 10^{12}\msun$ or higher
\citep{keres05,db06,cattaneo06,ocvirk08},
and the input becomes dominated by stars rather than gas.
Once the accretion cannot replenish
the gas discs on a time scale comparable to the timescale for the discs to turn
into clumps and stars, the galaxies gradually become dominated by old stars
and eventually stable against axi-symmetric modes. 
This is because, unlike the dissipating turbulent gas,
the stars maintain the velocity dispersion that they have acquired during
their history. While the young stars are expected to have a velocity
dispersion comparable to that of the gas and therefore to follow the gas 
into the bound clumps, the older stars are likely to have a higher $\sigma$. 
They therefore tend to form transient perturbations rather than accumulate
in bound clumps, and eventually they join the stable component that just
adds to the potential well and thus helps stabilizing the whole system.

The axi-symmetric instability of a two-component disc has been studied by
\citet{jog84} and \citet{rafikov01}. 
Denoting the velocity dispersions of stars and gas 
$\sigs$ and $\sigg$ respectively (with the latter standing for the speed of
sound if thermal pressure dominates), and defining $\Qs$ and $\Qg$ 
following \equ{Q} separately for each component, the effective $Q$ relevant
for the instability of the combined system is approximately
\be
Q^{-1} = 2\,\Qs^{-1} \frac{q}{1+q^2} 
            + 2\,\Qg^{-1} \frac{\siggs q}{1+\siggs^2 q^2} \, ,
\label{eq:Q2}
\ee
where $\siggs \equiv \sigg/\sigs$ and $q$ is the dimensionless wave number
$q \equiv k \sigs/\kappa$.
The first term has to be slightly modified to take into account the
dissipationless nature of the stars
but this correction makes only a small difference \citep{rafikov01}.
The system is unstable once $Q<1$. 
The most unstable wavelength corresponds to the $q$ that minimizes $Q$;
it lies between $q=1$ for $\siggs =1$ and $q \simeq \siggs^{-1}$ for
$\siggs \ll 1$.
Note that with $\sigs>\sigg$, the stellar disc by itself may tend to 
be less unstable than the gas disc by itself, $\Qs>\Qg$, but through its
contribution to the self-gravity that drives the instability, the stellar disc
can help the gas component de-stabilize the disc. The combined system can
be unstable even when each of the components has a $Q$ value above unity.
If, for example, $\Sigg=\Sigs$ and $\siggs=0.5$, the error made in $\Qc$ by 
ignoring the higher velocity dispersion of the stars is about 30\%.
If, on the other hand, $\siggs \ll 1$, the gas disc has to be treated on its
own while the stars become part of the stabilizing component.

The typical low-redshift discs are thus very different from the 
clumpy high-redshift discs.
In a bulge-dominated, cold disc, supernova and stellar feedback
may have an additional effect on the stabilization of the gas disc.
The common low-redshift discs are expected to form predominantly in haloes 
below the 
threshold mass of $\sim 10^{12}\msun$, and not necessarily by narrow streams
\citep{bd03,binney04,keres05,db06,bdn07}.
Being gas poor, the low-redshift discs rarely develop axi-symmetric 
instabilities with $Q<1$.
They can evolve secularly through non-axi-symmetric modes of instability with
$Q \gsim 1$, associated with quiescent star formation.
The turbulent high-redshift discs do not transform into today's thin
discs. They end up most naturally in today's thick discs
and S0 galaxies, or in ellipticals through mergers.

The above predictions of the simplified theoretical analysis
are to be tested below using cosmological simulations. In parallel, 
the instability analysis will help us interpret the simulated results.

\section{The Simulations}
\label{sec:models}

\begin{table} 
\caption{Simulations details}
 \begin{center}       
 \begin{tabular}{|lllcccc} \hline    
Comoving box size & 28.57~Mpc \\
Number of DM particles & 7 $\times 10^6$ \\
DM mass resolution  &  $5.5\times 10^5M_{\odot}$ \\
Max. resolution (proper)  &  35-70~pc \\
Min. mass  of a stellar particle  &  $10^4\Msun$ \\ 
 \end{tabular} 
 \end{center}
 \label{table:1}
 \end{table}

The simulations that serve as the basis for the current analysis follow the 
evolution of three typical massive galaxies.
They were performed completely independently, not tailored in any way to 
match the theoretical expectations or the observed SFGs.  
We selected dark-matter haloes with virial masses $\Mv \simeq 10^{12}\msun$ at
$z=1$, which end up as $(3-4)\times
10^{12}\msun$ today, comparable to and somewhat more massive than
the Milky Way.
Using the EPS statistic of halo growth \citep{Neistein06}, 
the mean virial mass of the major progenitor of such haloes
is expected to be $5 \times 10^{11}\msun$ at $z \sim 2.3$. 
The average input rate of gas into such a halo, assuming a universal baryonic
fraction of $\fb=0.165$, is 
$\Mbd \simeq 6.6\, M_{12}^{1.15} (1+z)^{2.25} \sy$,
or $\Mbd \simeq 44 \sy$ for the haloes in question.
These galaxies are thus at the low-mass end of the massive star-forming 
galaxies in the current surveys, in which the typical halo mass is above 
$10^{12}\msun$ and the typical SFR is $\sim 70 \sy$ 
\citep{Forster09}. This slight mismatch occurred because these
simulations were not originally intended for the particular study reported 
in the current paper.

The procedure for setting up the initial conditions of the simulations 
is as follows. We started with a low-resolution cosmological $N$-body 
simulation in a comoving box of side $20\hmpc$,
using the following cosmological 
parameters suggested by the WMAP5 results \citep{WMAP5}: 
$\omm=0.27$, $\oml=0.73$, $\omb= 0.045$, $h=0.7$ and $\sigma_8=0.82$.
At $z=1$, we selected three haloes of $\Mv \simeq 10^{12}\msun$ each,
termed \tres, \uno\ and \dos. 
The selection was arbitrary, except that we deliberately 
avoided haloes that were involved in a major merger process at $z \sim 1$. 
This selection rejected less than 20\% of the haloes of that mass
and it had no explicit dependence on the merger history prior to z=1.
One halo (\dos) happened to be in a relatively dense environment of neighboring
haloes, and the other two (\tres\ and \uno) were in a ``field" environment,
namely part of a typical low-density filament of the cosmic web.
These three haloes can therefore be regarded as a small fair sample of the 
haloes of that mass at that epoch.

For each halo, we identified for re-simulation with high resolution the 
concentric sphere of a radius twice the virial radius, typically one comoving 
Mpc.
This sphere was traced back to its Lagrangian volume at the initial time,
$z=60$, 
where we applied a zoom-in technique \citep{Klypin01} to refine the
fluctuations down to the desired resolution limit. Gas was added to the box
following the dark matter distribution with the universal
baryonic fraction $\fb=0.165$. 
We then re-simulated the whole box, with refined resolution only in the 
selected Lagrangian volume about the galaxy.

Each simulation has a total of $7\times 10^6$ dark-matter particles,
with four different masses. The high-resolution region 
is resolved with $\sim 4\times10^6$ dark-matter particles, each of mass
$5.5\times 10^5\msun$. 
The particles representing stars have a minimum mass of $10^4 \ \msun$ 
and their number increases as the simulation evolves; it reaches a  
value of $2-3 \times 10^6$ stellar particles at $z \simeq 1.3$. 
The integration of the gravity and the gas physics
are performed on an adaptive mesh, where
the maximum resolution is 35-70 pc in physical units at all times.
This resolution is valid, for example, throughout the cold disc 
and dense clumps.
The numbers characterizing each simulation are summarized in
table~\ref{table:1}.

The simulations were performed using the ART code
\citep{Kravtsov97, Kravtsov03}, following the evolution of the gravitating N-body system and the Eulerian
gas dynamics using an Adaptive Refinement Tree. 
The code incorporates the standard physical processes relevant 
for galaxy formation, as described in \citet{Ceverino09}.
These include gas cooling by 
atomic Hydrogen and Helium, metals, and molecular Hydrogen,
photoionization heating by a UV background with proper self-shielding, 
star formation, metal enrichment,
stellar mass loss,
 and stellar feedback.

In our stochastic star-formation model, more than $90\%$ of the stars form 
at temperatures well below $10^3$K and more than half the stars form below 
300~K at gas densities higher than $10 \cmc$.
New stellar particles are generated at every time step of the zero-level grid,
which is about 5 Myr at $z=2$.
We use a ``constant" feedback model, in which the combined energy from stellar 
winds and supernova explosions is released as a constant heating rate over 
40~Myr, the typical age of the lightest star that explodes as a type-II 
supernova.
The heating rate due to feedback may or may not overcome the cooling rate, 
depending on the gas
conditions in the star-forming regions \citep{ds86, Ceverino09}. 
We also include the effect of runaway stars
by assigning a velocity kick of $\sim 10 \kms$ to 30\% of the
newly formed stellar particles.
As a result, these stars can migrate $\sim100 \pc$ away from the dense regions 
and explode as supernovae in regions with lower densities and longer cooling 
times. This enhances the efficiency of feedback in heating the surrounding gas.
Finally, in order to mimic the self-shielding of galactic 
neutral hydrogen from the cosmological UV background, 
we assume that in regions where the gas density is higher than $n =0.1\cmc$,
the ionizing flux shrinks to a negligible value (i.e., 
$5.9 \times 10^{26} \ {\rm erg} \ {\rm s}^{-1} \ {\rm cm}^{-2} 
\ {\rm Hz}^{-1}$, the value of the pre-reionization UV background at $z=8$).

\section{Global Properties at High Redshift}
\label{sec:global}

\begin{figure}
\vskip 0.1cm
\center
\includegraphics[width =0.48\textwidth]{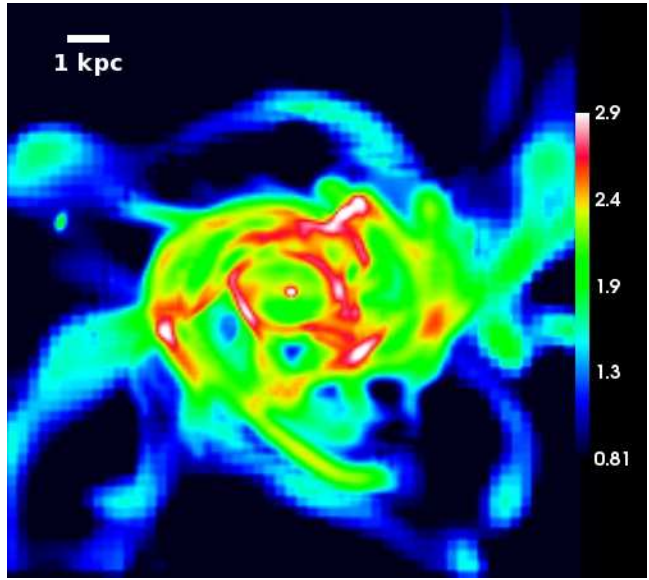}
\caption{Face-on gas surface density in galaxy \dos\ at $z=2.1$. The image
demonstrates violent disc fragmentation into transient features and bound
clumps, resembling observed SFGs and theoretical expectations.
The size of the image is $15\times 15\kpc$. Surface density is in units of
log($\msun {\rm pc}^{-2}$). For comparison, the surface density of
molecular clouds in low-redshift galaxies is $\sim 100\ \msun {\rm pc}^{-2}.$}
\label{fig:cover}
\end{figure}

Our analysis here focuses on the three simulated galaxies at
$z \sim 2-2.5$, when the Universe was roughly 3 Gyr old.
These redshifts are in the middle of the redshift range $1.5-3$ 
where the activity of disc instability and star formation is expected to be 
at its peak, both according to observations of SFGs and the predictions of DSC.
Figure~\ref{fig:cover} presents the gas surface density in galaxy \dos\ at
$z=2.1$. The face-on view angle is defined by the angular momentum of 
the $T<10^4$K gas inside a sphere of $3 \kpc$.
The image shows an extended disc of diameter $\sim 9 \kpc$
and the 70~pc resolution reveals that the disc is highly perturbed,
with a few big, round clumps embedded in massive elongated tangential features.
The elongated structures are transient features being disrupted by shear.
The clumps are bound, with typical masses $\sim 10^8 \msun$ and 
characteristic sizes $\sim 1 \kpc$.
About $10-20\%$ of the total mass of the disc is in the clumps and sheared 
features.
This galaxy, like the other two,
also has a stellar bulge, of mass comparable to the overall disc mass.
The very clumpy appearance of this high-redshift disc and its massive bulge,
which repeats in the other two simulated galaxies, 
is revealed for the first time in cosmological simulations.
A first visual inspection indicates that it resembles the clumpy appearance 
of the typical observed SFGs, and follows the theoretical expectations 
laid out by DSC, \se{theory}.

Before we proceed to analyze the detailed properties of the simulated
clumpy discs and bulges in comparison with the observed SFGs and the
theoretical predictions, we address the main global properties
of the dark-matter haloes and the stellar components in the simulated galaxies. 
We wish to verify in particular that they are consistent with the observed 
global properties of high-redshift star-forming galaxies, and their main 
scaling relations.

\subsection{Virial Properties of the Dark-Matter Haloes}
\label{sec:virial}

\begin{table}
\caption{Global properties of the host haloes at redshift $z=2.3$.
Radii are expressed in proper kpc, masses in units of $10^{11} \msun$
and velocities in $\kms$.}
 \begin{center}
 \begin{tabular}{cccc} \hline
\multicolumn{2}{c} {Galaxy}  R$_{\rm v}$ & M$_{\rm v}$ & V$_{\rm v}$  \\\hline
\tres \ & \ \ \ \ 70 & 4 & 150 \\
\uno \ & \ \ \ \ 68 & 3.5 & 140 \\
\dos  \ & \ \ \ \ 83 & 6.1 & 180 \\
 \end{tabular}
 \end{center}
\label{table:1b}
 \end{table}

Table~\ref{table:1b} displays the virial properties of the simulated haloes
at $z=2.3$. The virial radius $\Rv$ and the corresponding mass $\Mv$ are
defined to encompass a mean mass density of 180 times the universal mean, and
the virial velocity is $\Vv \equiv (G \Mv/\Rv)^{1/2}$.
Haloes \tres\ and \uno\ are similar, with $\Mv \simeq 4 \times 10^{11} \Msun$
and $\Vv \simeq 150 \kms$, while halo \dos\ is a bit more massive with
$\Mv =6 \times 10^{11} \msun$ and $\Vv=180\kms$.

Galaxy \dos\ has a growth history that differs from the other two.
By $z=2.3$, its halo has already assembled about half the mass that it
will have at $z=1$,
while haloes \tres\ and \uno\ have assembled only a quarter of their $z=1$
mass.
The growth of \dos\ is thus faster prior to $z=2.3$, and the growth of the
other two is more rapid after $z=2.3$.
This reflects the fact that halo $\dos$ was forming in an environment of higher
density.
The cosmological overdensity of matter in a concentric sphere of comoving
radius $5\hmpc$ about haloes \tres\ and \uno\ at $z=2.3$ is only
$\delta\rho/\rho \simeq 0.07$ and $0.02$ respectively,
while it is as high as $\delta\rho/\rho \simeq 0.3$ for halo \dos.
This difference, representing the ``cosmic variance", was set by our selection
of haloes.
We will see that the cosmic environment plays an important role in the
formation of galaxies at the centres of these haloes, especially because
the earlier formation time of halo \dos\ makes it denser in its central
regions, both in dark matter and in gas.

\subsection{Scaling Relations at High Redshift}
\label{sec:scaling}

\begin{figure}
\center
\includegraphics[width =0.45 \textwidth]{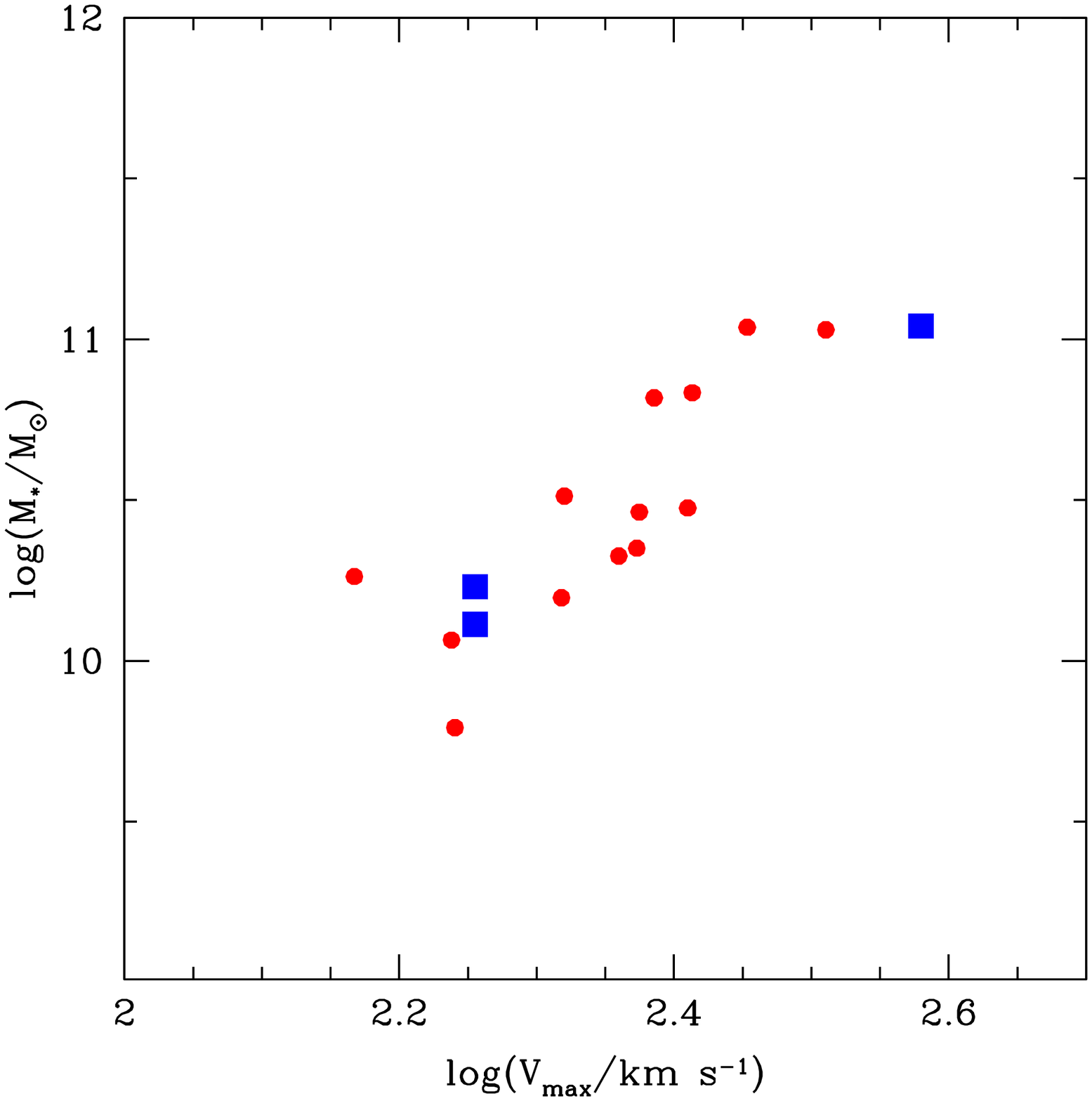}
\includegraphics[width =0.45 \textwidth]{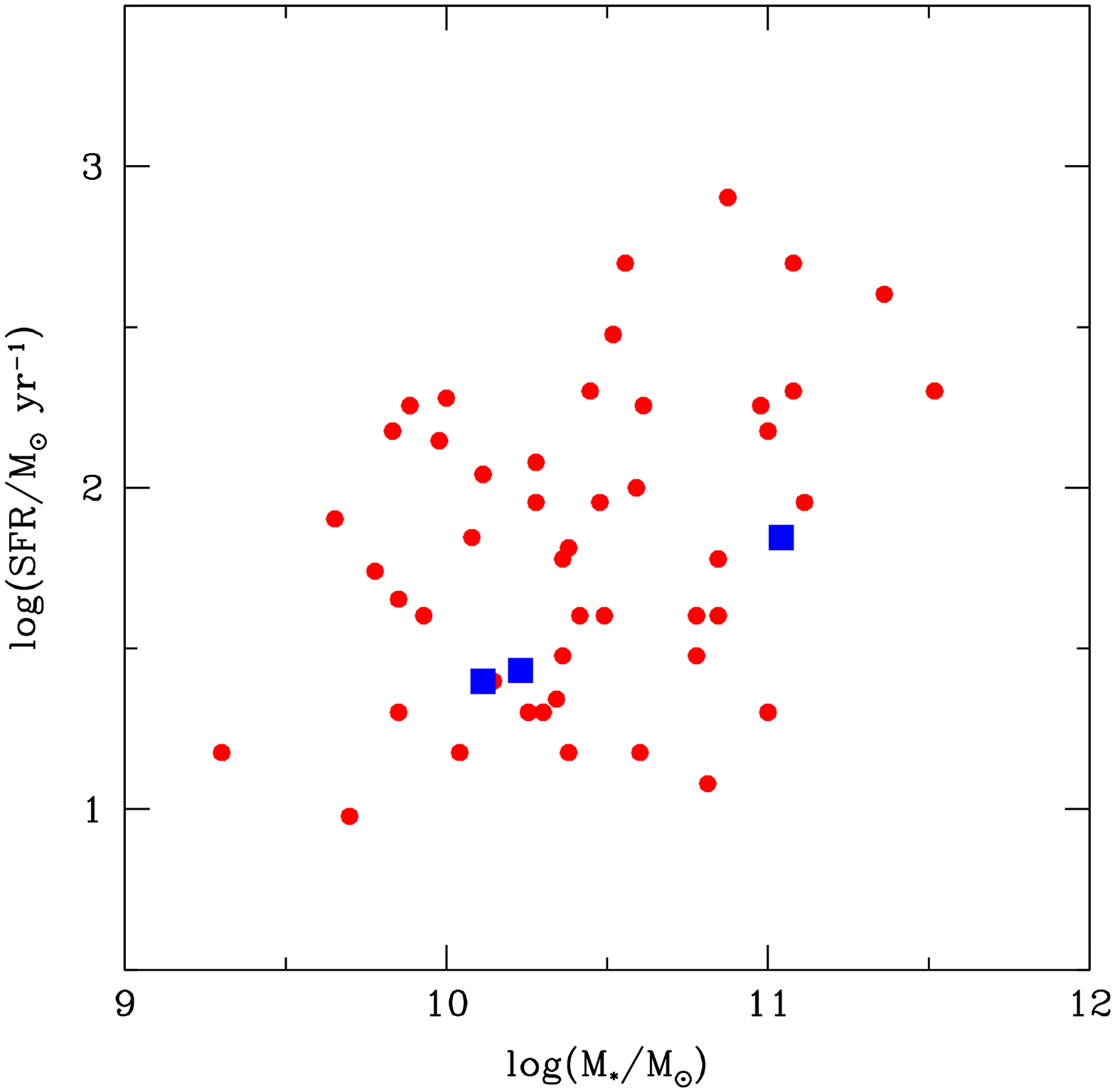}
\caption{
Scaling relations at $z \sim 2.3$ for our three simulated galaxies (blue squares) in comparison to observed galaxies (red circles). Observations are from the SINS survey \citep{Forster09,  Cresci09}. The top panel shows the Tully-Fisher relation, stellar mass versus maximum rotation velocity.
The bottom panel shows the ``main sequence" of star-forming galaxies,
SFR versus stellar mass.
The simulated galaxies seem to obey the observed scaling relations
within the given scatter.
They lie near the best-fit line of the Tully-Fisher relation,
but they have SFR values about a factor of two lower than the average value
for their stellar mass.}
\label{fig:sinfoni}
\end{figure}

\begin{figure*}
\includegraphics[width =\textwidth]{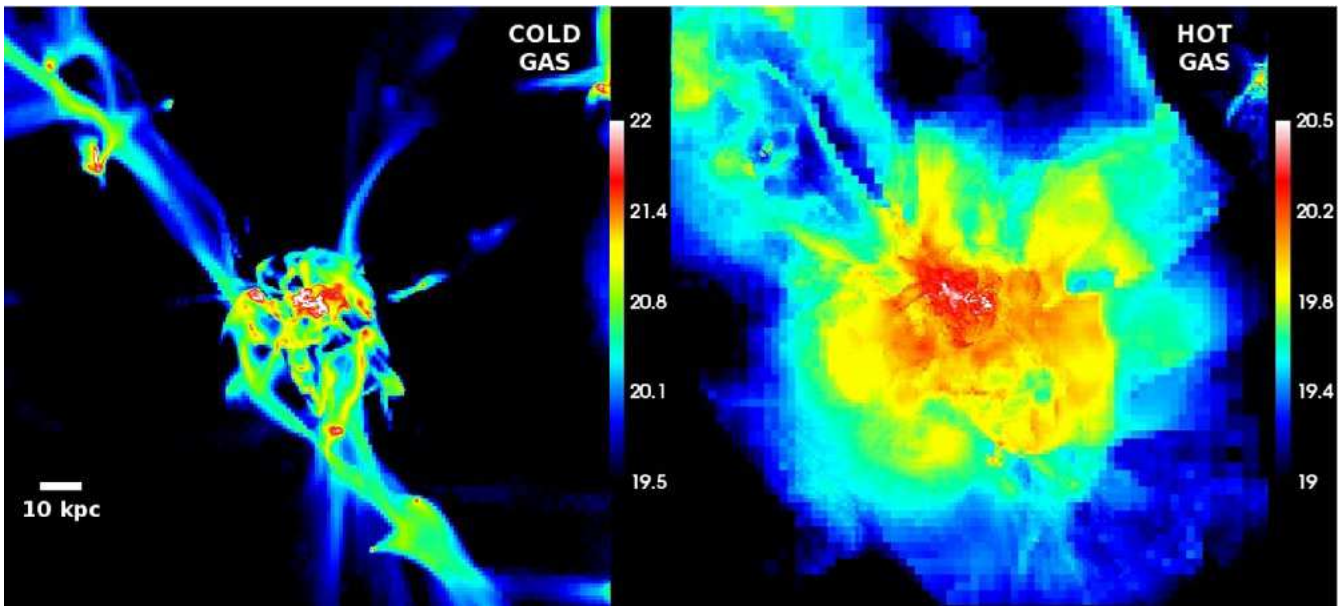}
\caption{
A zoom-out gas surface-density maps showing the streams feeding
galaxy \tres. Left: cold gas ($T<5\times 10^{4}$ K). 
Right: hot gas ($T>3\times 10^{5}$ K).
The box size is $160\times 160\kpc$, covering the whole virial sphere.
The color refers to log gas surface density in units of H atoms cm$^{-2}$.
Two major narrow streams carry the gas from well outside the virial radius
to the inner $\sim 20 \kpc$ halo core, where they break into a multi-stream
turbulent core before joining the inner disc of radius $\sim 6 \kpc$, 
seen nearly edge-on at the box centre (mostly in white). 
}
\label{fig:zoomout}
\end{figure*}

Figure~\ref{fig:sinfoni} refers to the two main scaling relations of SFGs, 
correlating the total stellar mass to SFR and to disc rotation velocity.
The observational data is extracted from the SINS survey
\citep{Forster09} --- a spectroscopic imaging survey
in the near-infrared with SINFONI/VLT that focuses on SFGs at $z\sim2$. 

The top panel of figure \ref{fig:sinfoni} puts our simulated galaxies
in the context of the Tully-Fisher relation. The observational data
is taken form the SINS sample of disc galaxies \citep{Cresci09}.
Since the rotation curves in the simulated galaxies are rather flat outside
the bulge, we adopt the average rotation velocity in the gaseous disc as our estimate
for $V_{\rm max}$ (see \se{GalProp}).
The simulated galaxies seem to follow quite well
the observed Tully-Fisher relation.

The bottom panel addresses the ``main sequence" of star formation, namely the 
SFR as a function of total stellar mass.
Given the scatter, the three simulated galaxies are consistent with
the observed relation. The fact that they lie a factor of $\sim 2$ 
below the mean SFR at a given mass may be a random coincidence or it may
reflect a certain deficiency in the simulated star-formation history.
The stellar masses in \tres\ and \uno\ are at the level of 5\% of
their halo masses of $\sim 4 \times 10^{11}\msun$, which is quite typical for
such haloes. The SFR in these galaxies are comparable to and only slightly
smaller than the expected average gas accretion rate into haloes of similar
masses, $\sim 30 \sy$.
In galaxy \dos, with a halo only slightly more massive but a significantly
denser environment and an earlier growth, the stellar mass
is higher than average, and the SFR is higher than the average accretion rate.

The agreement with the Tully-Fisher relation
may indicate that the small systematic deviation in the SFR-$M_*$ relation
reflects a deficiency in the current SFR at $z=2.3$ rather than in stellar mass.
It is thus possible that the simulated galaxies overproduce stars prior
to $z \sim 2.3$, and then underestimate the gas fraction and the SFR at 
$z \sim 2.3$, while they reproduce the proper stellar mass.
This may result, for example, from slightly underestimating the effect of 
supernova feedback at high redshifts.
Indeed, the SFR, and how it is affected by stellar and supernova feedback,
are naturally the most uncertain elements in our simulations, as these
processes are implemented via crude recipes of subgrid physics.
Despite this potential small deficiency, we conclude that the simulated
galaxies do obey in a satisfactory way the main observed scaling relations,
so their global properties can be considered as representative of typical
SFGs of a similar mass at a similar time.

\section{Clumpy Discs and Bulges in Steady State}
\label{sec:results}

\subsection{Feeding by Cold Streams}

Figure~\ref{fig:zoomout} shows a virial-scale view of the gas surface-density 
in halo \tres.  The cold gas, $T<5\times 10^4$K, 
flows along narrow streams that come from outside the virial radius
and penetrate to the centre of the dark-matter halo. There, they blend
in the inner $\sim 20 \kpc$ into a turbulent region 
and eventually settle in a clumpy disc of radius $\sim 6 \kpc$. 
The shock-heated gas of $T>3\times 10^5$K fills much of the virial sphere.
In the disc vicinity, the hot gas is also clumpy, perhaps
indicating the presence of shocks at the interfaces between the streams
and the central galaxy.

\subsection{Images of Gas, Stars and Dark Matter}
\label{sec:images}

\begin{figure*}
\includegraphics[width =0.68\textwidth]{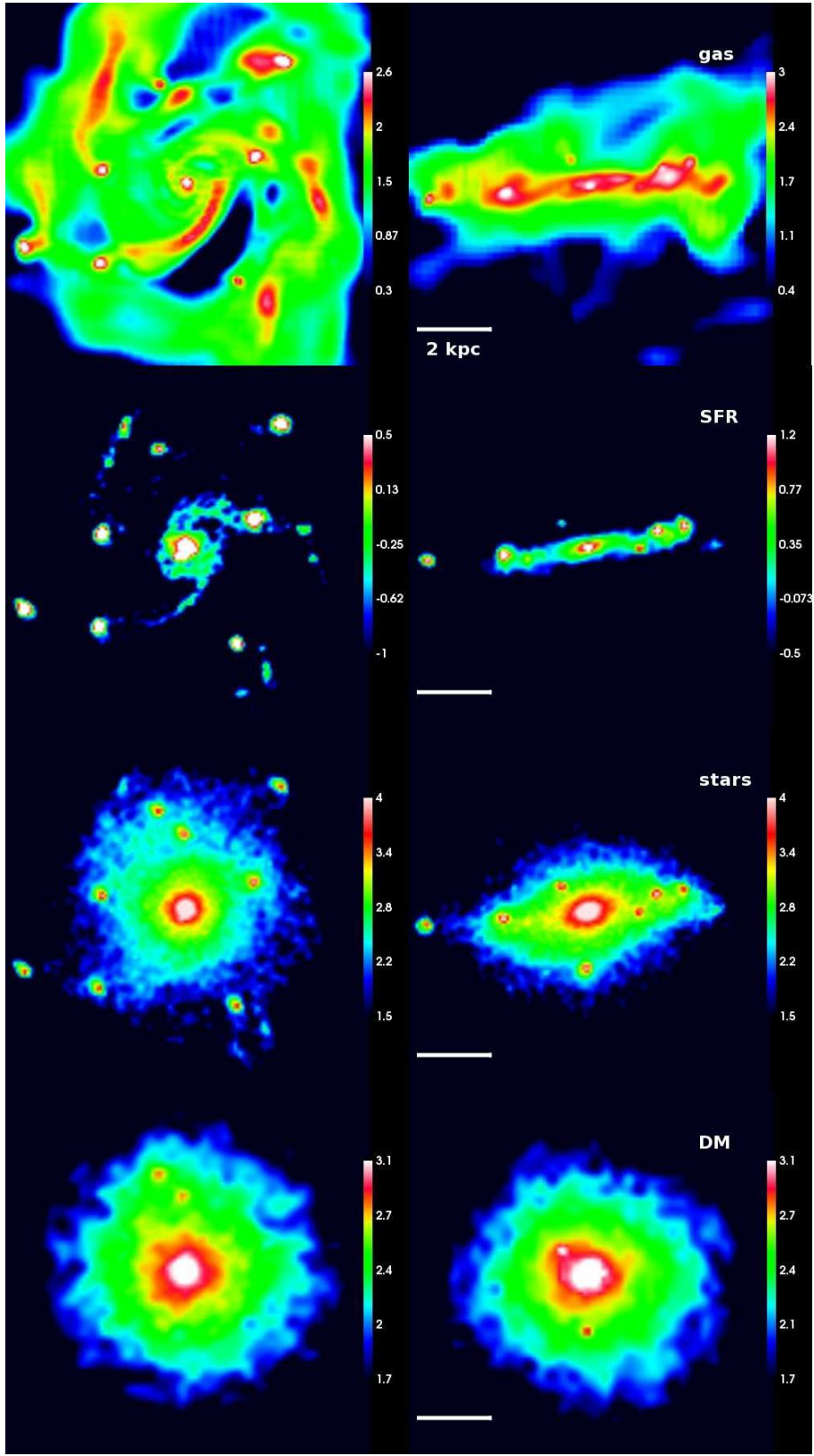}
 \caption{Surface density maps of galaxy \tres\ at $z=2.3$.
The side of each panel is $10\kpc$.
Shown are face-on views (left) and edge-on views (right).
From top to bottom: gas, SFR, stars, and dark matter.
The units are log($\msun {\rm pc}^{-2}$) for the surface density
and log($\sy {\rm kpc}^{-2}$) for the SFR. 
We see a disc of gas and young stars, a significant bulge extending to
a very thick disc of old stars, and a rather spherical
dark-matter halo. The disc is fragmented to elongated features
and round dense clumps. Stars form at a high rate in the clumps
and at the centre of the bulge. Most of the clumps are in-situ to the disc
and are not associated with dark-matter concentrations.
Two of the clumps are off-disc satellites. 
The color code was chosen to emphasize the clumps at the expense of
saturating the highest densities.}
\label{fig:1}
\end{figure*}

\begin{figure*}
\includegraphics[width =0.69\textwidth]{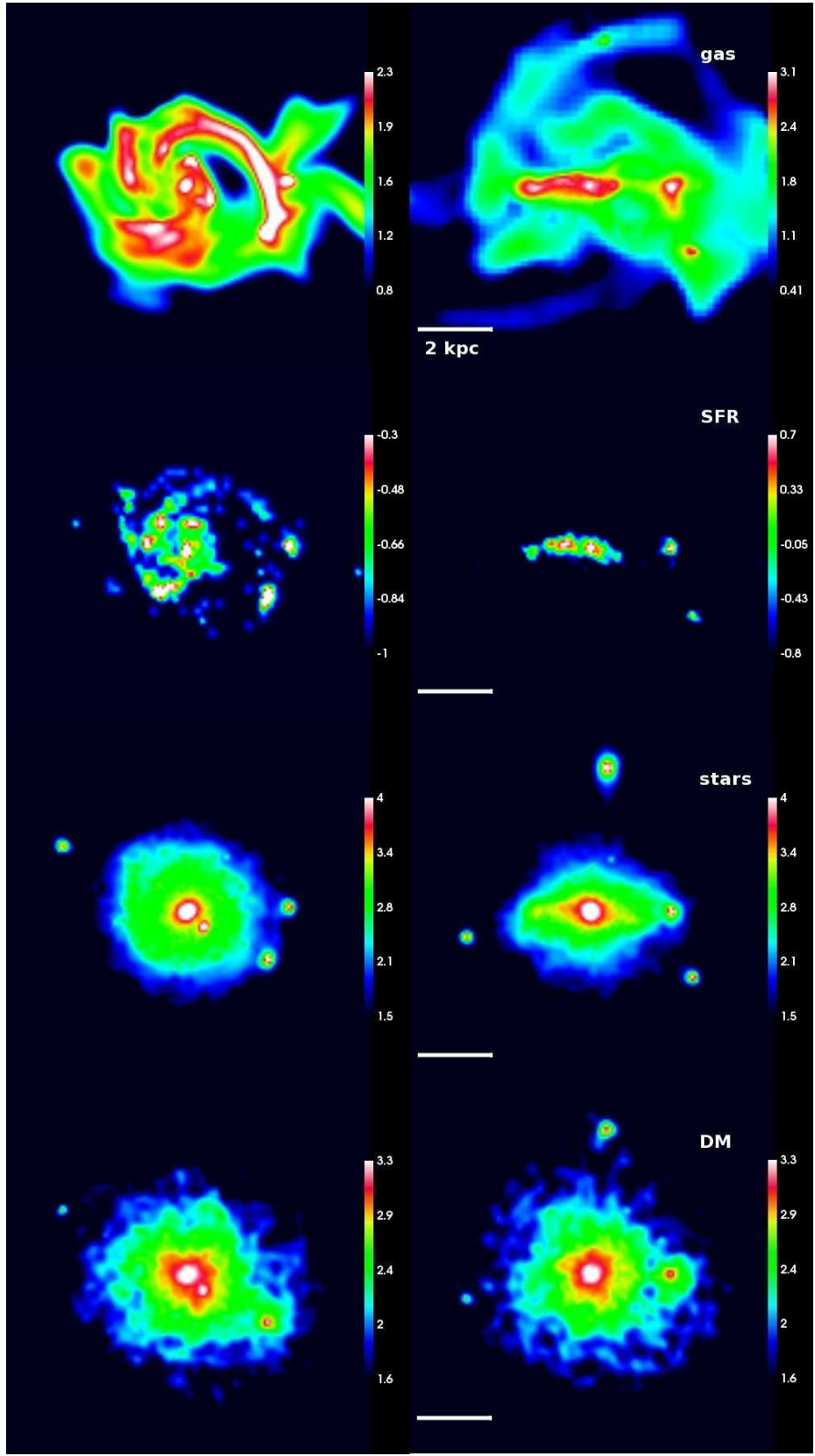}
 \caption{Same as in figure \ref{fig:1} but for Galaxy \uno. 
This case shows the smallest disc of the sample. 
Beyond the high SFR in clumps, there is a relatively high SFR in other 
regions spread across the disc, some associated with the elongated transient 
features.
The overall stellar population is rather smooth and concentrated in the bulge.
Two clumps are external satellites with stars and dark matter.
}
\label{fig:2}
\end{figure*}

\begin{figure*}
\includegraphics[width =0.69\textwidth]{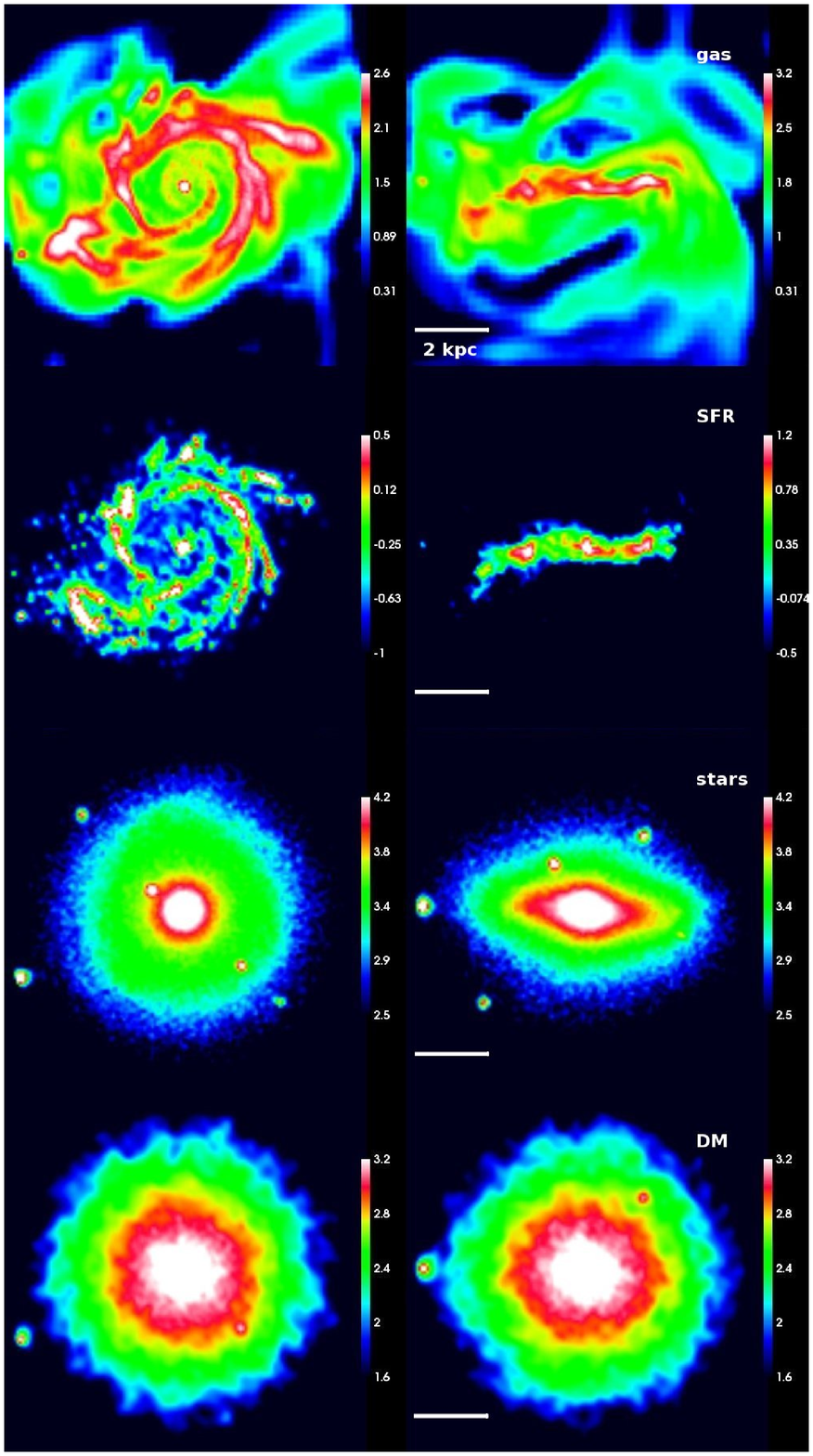}
 \caption{Same as in figure \ref{fig:1} but for Galaxy \dos. 
This is the most massive and early forming galaxy in our sample. 
It has the highest surface densities of gas and stars and
is therefore highly unstable. 
Star formation occurs in an extended clumpy ring.
Two clumps are external, with stars and dark matter.
}
\label{fig:3}
\end{figure*}

Figures~\ref{fig:1}-\ref{fig:3} show face-on and edge-on views of the 
galaxies in the inner $\sim$10\% radius of the three haloes at $z=2.3$. 
The maps display on the same scale the distribution of gas, SFR, stars and 
dark matter.

Figure~\ref{fig:1} refers to galaxy \tres. 
The gas surface density maps show a thick disc of radius $\sim6\kpc$ 
and thickness $\sim 1\kpc$. The disc is highly fragmented showing several 
giant clumps and elongated transient clouds as in Figure \ref{fig:cover}.
The SFR, as deduced from the stars younger than $10\Myr$, 
reveals that 
a large fraction of the stars form
in the disc clumps and in the centre of the bulge. 
The stars show a thick disc with embedded clumps, all associated with the gas,
star-forming clumps, but the stellar distribution is dominated by a massive
bulge, 
of a mass comparable to the disc mass.
The bulge extends to a very thick disc,
perhaps resembling a lenticular galaxy.
The dark-matter distribution is rather spherical and smooth, with only two 
obvious dark-matter satellites, both off the disc plane. 
The disc clumps are thus all naked, not associated with small dark-matter 
haloes of their own.  This is evidence for the in-situ nature
of the disc clumps, indicating that they were originated in the disc itself
rather than merged as small galaxies coming from the outside 
along the streams.
The two off-plane satellite galaxies have 
haloes of masses $10^8-10^9\ \msun$ and
eccentric orbits that lead them to future mergers 
with the bulge while they never become part of the rotating disc.

Figure~\ref{fig:2} shows galaxy \uno. This galaxy, in the smallest halo of the
three, has a disc radius of only $\sim 3.5\kpc$. The gaseous disc is strongly 
perturbed, with clumps embedded in elongated massive features that resemble
spiral arms.  The gas clumps here are less compact than in galaxy 
\tres, with a lower surface density extending to larger radii.
The low stellar fraction in these clumps, 
compared to the clumps of galaxy
\tres, reflects the transient nature of these features.
The overall SFR of $\sim30 \sy$ is similar to galaxy \tres\ but it is spread 
over the clumps and the broad elongated features covering a large fraction
of the disc.
The dark-matter map shows two satellites, coinciding with two clumps of gas 
and stars, of which one is actually in the disc plane. All the other clumps
and elongated features are in-situ to the disc and not associated with
dark-matter substructures.  
A non-negligible stellar bulge is seen, 
with a mass comparable to the disc mass,
but it is less massive than in the other two cases.

Figure~\ref{fig:3} displays galaxy \dos, the most massive case that assembled 
most of its dark matter and accreted most of its baryons earlier than the other
two. 
The gas disc is very clumpy and disturbed, despite the relatively low gas
fraction of less than 10\% in this galaxy. The instability is largely because 
the disc is more massive and more compact than in the other cases, with 
a high surface density in both gas and stars,
$\sim2 \times 10^2$ and $3 \times 10^3 \msun {\rm pc}^{-2}$ respectively.
The clumps and elongated features display a star-forming ring outside the 
bulge.  
This structure is remarkably similar to observations of star-forming outer 
rings in some $z \sim 2$ galaxies \citep{Genzel08}.
Two off-plane satellites are seen in the stars, of which at least one is 
clearly associated with a dark matter satellite. The main features in the disc
are not associated with dark-matter substructure.
As in the other two cases, the bulge is as massive as the disc.
 
In general, all three cases show a gravitationally unstable, extended,
thick and irregular disc, with massive in-situ clumps and transient features, 
accompanied by a compact bulge of comparable mass, 
consistent with 
observations and the cosmological steady state predicted by theory.
The mean gas surface density within the clumps is high, 
$\Sigg\sim500-1000 \ \msun {\rm pc}^{-2}$, compared to the typical values 
of $\sim 100 \ \msun {\rm pc}^{-2}$ in the ``giant" molecular clouds in 
low-redshift galaxies \citep{Bolatto08,Heyer08}.
This allows high, starburst-like, SFR 
surface densities of $\sim 10 \sy \ {\rm kpc}^{-2}$, as expected from the 
Schmidt-Kennicutt relation \citep{Kennicutt98}.
Overall, $10-20\%$ of the gas in the discs is in the form of clumps
and sheared features.
They are larger and much more massive than the
star-forming molecular complexes in quiescent, low-redshift disc galaxies.
This makes the gravitational interaction between the clumps and the
transient features more intense,
resulting in high turbulence and rapid mass flow to the center.
The surface density in the transient clouds is $(100-500) \msun {\rm pc}^{-2}$ 
and they are embedded in a smoother inter-clump medium of
surface density $\sim$30 $\msun {\rm pc}^{-2}$.

\subsection{A Preliminary Comparison with Observed Images}
\label{sec:UDF}

\begin{figure*}
\includegraphics[width =0.69\textwidth]{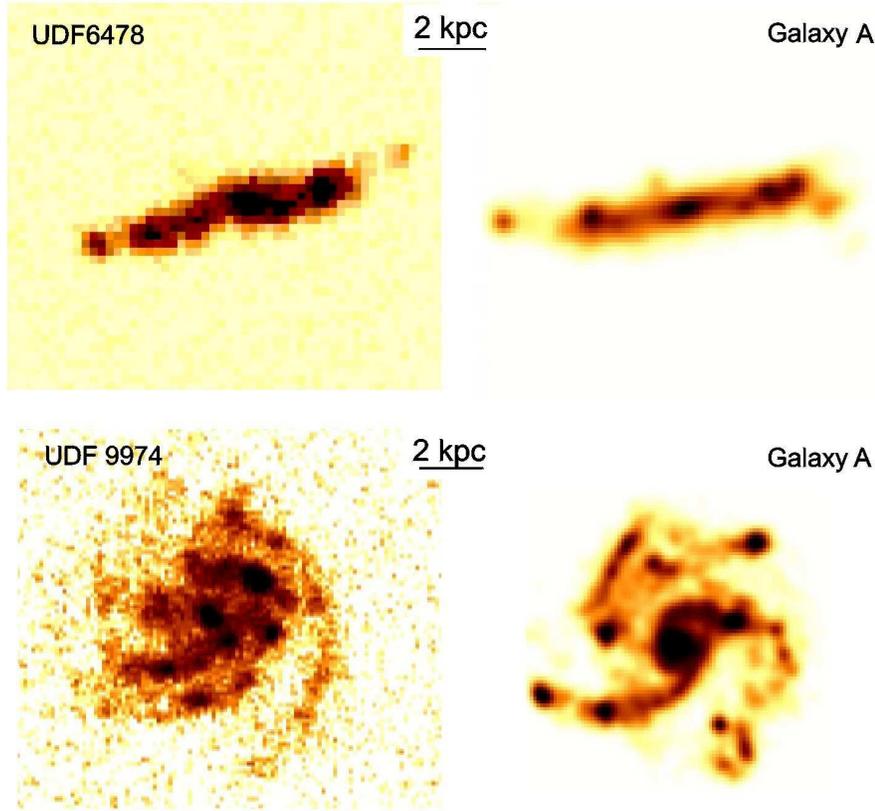}
\caption{
Typical images of observed and simulated clumpy disks at high redshifts.
Left: UV rest-frame surface brightness of two galaxies from the Hubble Ultra
Deep Field, representing the different appearances of the clumpy discs
depending on the viewing angle: a ``chain" galaxy (UDF~6478, z-band
at $z=3.3$) and a face-on ``clump cluster'' (UDF~9759, V-band at $z=1.2$).
The 3$\sigma$ sensitivity in the images is 25.5 mag arsec$^{-2}$ and 
the surface brightness of the brightest clump (in black on the images) is around 21 mag arcsec$^{-2}$.
 Right: The simulated galaxy A viewed at  $z=2.3$ with about the same resolution, using as
a proxy for the UV emission the stars younger than 100~Myr.
The maximum and minimum surface densities are 0 and 4 in log($\msun {\rm pc}^{-2}$).
The color code is logarithmic and the same for the observations and the simulations.
The pixel size and seeing are assumed to be 0.11 and 0.5 kpc, respectively.
We do not aim here at modeling individual galaxies or making a valid statistical
comparison between the simulated and observed morphology. We pick the
observed galaxies at redshifts that bracket our fiducial simulation at $z=2.3,$ 
and because there is observational evidence for
a similar morphology in the redshift range $z=1-4$ \citep[e.g.,][]{Elmegreen07}.
A visual comparison indicates that our simulations may reproduce the key morphological
properties of high-redshift clumpy galaxies.}
\label{fig:sim_obs}
\end{figure*}k

In \fig{sim_obs} we show side by side images of simulated and observed galaxies,
as a sneak preview for a quantitative comparison to be provided
elsewhere (Ceverino, Bournaud \& Dekel 2009, in prep.).
Edge-on and face-on images of simulated galaxy A at $z=2.3$ are shown next to a ``chain" and a ``clump-cluster"
galaxy respectively.
Since the young stars contribute a significant fraction of the rest-frame
U-band luminosity, the images of simulated stars younger than 100 Myr
are compared to observed UV rest-frame images from the Hubble Ultra Deep
Field (HUDF).
The mean rest-frame UV (observed i band at z=2.3) mass-to-light ratio of the simulated stellar population is 0.1 $ \msun/\lsun,$ according to GALAXEV population synthesis code \citep{BC03}
The pixel size and seeing are assumed to be 0.11 and 0.5 kpc, respectively.
Galaxy A is shown at $z=2.3$ because this is near where the clumpy appearance peaks in this particular simulation.
The chosen UDF galaxies are at $z=3.3$ and $z=1.3$ respectively and they bracket our fiducial model at  $z=2.3.$
This is fine for a very qualitative visual comparison given the observational evidence for a similar clumpy morphology
of observed star-forming galaxies across the redshift range $z=1-4$ \citep[e.g.,][]{Elmegreen07}.
The weak sensitivity to redshift is also consistent with the analytic prediction by DSC of a cosmological steady state.
Therefore, for this very qualitative visual comparison, we did not make an effort to fully match redshifts, luminosities, masses and other galaxy properties,
except for the overall similarity of disk size and the clumpy nature of the young stellar population.
A visual comparison indicates that the general morphological features of the
high-redshift clumpy SFGs are successfully reproduced by our simulations.
At least in the specific cases shown, the similarity includes the number and sizes
of clumps, their distribution throughout the system, and the contrast
between the clumps and the inter-clump medium.
This preliminary comparison is for illustrative purposes only --- it is 
neither meant to provide a detailed modeling of the two specific galaxies, 
nor to present a statistically significant comparison of fair samples.
 
The dominance of a few giant clumps is a common feature to the 
simulated and observed galaxies. For example, in galaxy \tres, about 40\% of 
the young stars are in the in-situ disk clumps, and 10\% in the bulge. 
The inter-clump medium, which is naturally more evident in the edge-on views, 
is the site for the other half of the SFR, which is associated with diffuse 
inter-clump UV light. This is consistent with the observation
that $\sim$30\% of the light in cluster galaxies is coming from clumps
\citep{Elmegreen05}.
The clump sizes, their contrast with the rest of the disc, and their
spatial distribution throughout the system, all seem strikingly  similar
between simulation and observation. In particular, some of the simulated
galaxies show asymmetric configurations, resembling the observed
``tadpole" galaxies.
The mean stellar age within individual clumps, averaged over all the clumps 
of galaxy \tres, is $\sim$80 Myr. The dispersion of stellar ages within a clump 
is comparable, $\sim$70 Myr. This implies that the star-formation history 
within each clump is continuous over a period of $100-200 \Myr$. 
These stellar ages are similar to the age estimations from observations 
\citep{Elmegreen09}.
The mean stellar ages within the bulge is $\sim$1 Gyr, with a
spread of $\sim$0.5 Gyr, in agreement with the observational estimates
\citep{Elmegreen09}.

All three simulated galaxies host central stellar bulges with  
masses comparable to the disc mass, in agreement with the estimated
bulge-to-total mass ratio from the SFGs at high redshift
\citep{Genzel08,Elmegreen09}.
These stellar spheroids are rather compact, with radii $(1-2)\kpc$,
compatible with the spheroids detected at high redshifts
\citep{Kriek06,Dokkum08,Bezanson09}.

\subsection{Galaxy Properties vs Theory and Observation}
\label{sec:GalProp}

\begin{table*} 
\caption{Galaxy properties at $z=2.3$. Radius in kpc and mass in
$10^{10}\msun$.}
 \begin{center}       
 \begin{tabular}{ccccccccccccc} \hline     
\multicolumn{2}{c} {Galaxy}  R$_{\rm disc}$ & R$_{\rm bulge}$ & M$_{\rm gas}$ & M$_{\rm stars}$ & M$_{\rm DM}$ & M$_{\rm bulge}$ & M$_{\rm disc}$ & M$_{\rm disc}/$M$_{\rm total}$ & M$_{\rm gas}/$M$_{\rm disc}$ & M$_{\rm bulge}$/M$_{\rm disc}$ & M$_{\rm clumps}/$M$_{\rm disc}$\\\hline 
\tres & 5.8 & 2.0 & 0.37 & 1.73 & 2.1      & 1.0   & 1.1    & 0.26 & 0.35 & 0.93  & 0.14 \\
\uno & 3.4 & 1.0 & 0.20 & 1.27 & 0.92   & 0.81 & 0.66  & 0.27 & 0.30 & 1.22 &  0.24 \\
\dos & 4.6 & 1.5 & 0.46 & 11.44  & 3.4      & 6.6   & 5.3    & 0.35 & 0.09  & 1.25 & 0.15 \\   
 \end{tabular} 
 \end{center}
\label{table:2}
 \end{table*}


Table \ref{table:2} lists global properties for the main components of
the three galaxies: cold gas (T$<10^5$ K), stars and dark matter inside 
the disc radius, $\Rd$.
This radius is estimated from a visual inspection of figures
\ref{fig:1}-\ref{fig:3} and the associated density profile of the stellar and 
gas disc, which are both truncated at a similar radius.
The bulge radius $\Rb$ is read quite straightforwardly from the stellar 
density profile, which shows a well defined transition from a steep inner 
profile characteristic of a bulge and a shallower outer exponential profile 
with an exponential radius of $1-2 \kpc$, comparable to the bulge radius.
The density profile of the gas disc is much shallower, with an exponential 
radius comparable to the outer disc radius $\Rd$.

We use the kinematics of the star particles to distinguish the stellar disc 
from the spheroidal component. We assign a star particle to the disc only if 
the $z$-component of its angular momentum $j_z$ (parallel to the total galaxy 
angular momentum) is higher than a fraction $\fj$ of the maximum angular 
momentum for the same orbital energy, $j_{\rm max}=|v| \ r,$ 
where $|v|$ is the magnitude of the particle velocity and $r$ is its 
distance from the center.
The distribution of $j_z/j_{\rm max}$ among the simulated stars is  
bimodal, reflecting the division to distinct kinematic components of disc 
and bulge.
We adopt as default $\fj=0.7$, which roughly marks the transition   
between the two components of this distribution.
We thus define the bulge mass, $M_{\rm bulge}$, 
as the mass in stars inside $\Rd$ with $j_z/j_{\rm max}>\fj$, 
and the disc mass, $M_{\rm disc}$, as the mass of cold gas, $M_{\rm gas}$,
plus the mass in stars with $j_z/j_{\rm max}>\fj$.

We find in the three galaxies values of $\delta \sim 0.3$
for the disc-to-total mass ratio (\equ{delta}).
These are compatible with the values predicted by DSC for unstable discs 
in cosmological steady state. The simulations thus reveal the predicted  
even distribution of mass between the disc, bulge and dark-matter components 
within the disc radius.

The gas fractions in the discs of \tres\ and \uno\ are 35\% and 30\%
respectively. This is significantly higher than in today's typical spirals, 
where the gas fraction is typically $\sim 10\%$ or less.
It is comparable, and perhaps on the low side of the crude observational
estimates for typical high-redshift SFGs,
both the indirect estimates based on the Kennicutt-Schmidt relation
\citep{Erb06,Genzel06}
and the direct estimates based on CO observations
\citep{Daddi04,Greve05,Tacconi06,Tacconi08,Daddi08,Daddi09}. 
The gas fraction in galaxy \dos\ is lower than 10\%, making it atypical.
This may reflect an overproduction of stars too early in this simulation
(see \se{disc}).
The baryonic fractions within the disk radius in the three simulated galaxies,
0.5, 0,87, and 0.77,  
are consistent with the observational estimates \citep[][Fig.~20]{Forster09}.
The fact that the gas fractions are lower than expected 
should be taken into account in the
instability analysis, but it is encouraging to note that the clumpy
appearance of the discs, and even the values of $\delta$ as derived from
the combined disc of gas plus stars, are consistent with the observations 
and the theoretical predictions.

\begin{table} 
\caption{Star formation and baryonic fraction. 
SFR in $\sy$, star-forming gas mass in 10$^{10}$ $\msun$, and star-formation time in Myr.}
 \begin{center}       
 \begin{tabular}{ccccc} \hline     
\multicolumn{2}{c} {Galaxy}   SFR & M$_{\rm SF\_gas}$ & t$_{\rm SF}$  & M$_{\rm galaxy}$/M$_{\rm v}$ \\\hline 
\tres & 27 & 0.21  & 78 &   0.052  \\
\uno & 25 & 0.13  & 52 &   0.042  \\
\dos & 70 & 0.33  & 47  &   0.19  \\ 
 \end{tabular} 
 \end{center}
\label{table:2b}
 \end{table}

Assigning gas to clumps using a density threshold of $n = 25 \cmc$
(or any value in the range $(10-40)\cmc$), 
we obtain that the fraction of the disc in clumps is $0.15-0.25$. 
Indeed, these fractions in clumps 
are consistent with the value of $\alpha \sim 0.2$ 
adopted in the analysis of DSC.
The mean clump mass in all cases is $\sim 10^8 \msun$, in agreement with 
the prediction, \equ{Mc}. 
The clump radius ranges from 0.2 to 1 kpc, consistent with the pre-collapse 
radius predicted in \equ{Rc} for galaxies of such mass, $\Rc \sim 0.8 \kpc$.
The bound clumps, as opposed to the transient features, have rather 
sharp boundaries, and they are approximately in virial equilibrium.

 \begin{table} 
\caption{Kinematic disc properties. Velocities are in $\kms$.}
 \begin{center}       
 \begin{tabular}{ccccccccc} \hline     
\multicolumn{2}{c} {Galaxy}   $V_{\rm gas}$ & $\sigma_{\rm r, gas}$ & $(\sigma_{\rm r}/V)_{\rm gas}$  & $V_{\rm stars}$ & $\sigma_{\rm r, stars}$ & $(\sigma_{\rm r}/V)_{\rm stars}$ \\\hline 
\tres & 180 & 20  & 0.11 &   160 & 33 & 0.22 \\
\uno & 180 & 25  & 0.14 &   157 & 34 & 0.26 \\
\dos & 380 & 60  & 0.16  &   356 & 107 & 0.30 \\ 
 \end{tabular} 
 \end{center}
\label{table:3}
 \end{table}

Table \ref{table:2b} lists galaxy properties that are related to 
star formation (discussed in section \ref{sec:scaling}) and baryonic fraction.
The average SFR of 45 $\sy$ is in the ballpark of the expected mean gas 
accretion rates, \equ{Mdot}, with the SFR in galaxy \dos\ slightly above 
the average. This is consistent with the main finding of \citet{dekel09},
based on a large hydrodynamical cosmological simulation,
that a few cold streams carry almost all the accreted gas into the inner 
galaxy where it turns into stars in a rate comparable to the accretion rate.
In order to estimate the timescale for star formation, 
or for gas consumption, we compute the mass of ``star-forming gas",
$M_{\rm SF-gas}$, inside the disk radius $R_{\rm d}$. 
We define this as the gas with a density higher than 1 $\cmc$ and a 
temperature lower than $10^4$ K. It represents 60-70\% of all the gas in 
the disc. The ratio between the star-forming gas and the SFR defines a SFR
timescale of $t_{\rm SF}=50-80 \Myr$, which is somewhat shorter
than estimated from observed galaxies at high redshift \citep{Forster09}.

Two of the simulated galaxies show a low mass fraction of galactic baryons 
inside the dark-matter halo, which is 25-32\% of the cosmological baryonic 
fraction $\fb$, probably reflecting gas outflows from the small building 
blocks of these galaxies. These values are in accord with the baryonic 
fractions that are estimated for massive, present-day galaxies 
(F.~Prada, private communication).
The baryonic fraction in galaxy C is high, and it
may reflect an exceptionally high baryonic accretion-rate history.

Table~\ref{table:3} summarizes the global kinematics properties of the gaseous
and stellar disc: the rotational velocity, $V$, and the radial velocity
dispersion, $\sigma_r$. These quantities are the density-weighted mean values
for the disc outside the bulge, $r>\Rb$, 
where the rotation curve is roughly flat.  The stellar disc is defined 
using the kinematic method described above.
Galaxies \tres\ and \uno\ show similar overall kinematics. 
Their rotational velocity of $V\simeq 160 \kms$ 
is comparable to that of today's discs in similar haloes. However, their 
velocity dispersion of $\sigma_r \simeq 30 \kms$ is higher than in 
today's discs, reflecting a thick and irregular disc in which the pressure 
is dominated by turbulence. 
Within the clumps, the internal gas velocity dispersion is $\sim 20-30 \
\kms$, and here too it is the dominant source of pressure.
Galaxy \dos\ is again different; 
its rotational velocity and dispersion are significantly higher than in 
the other two cases.
The $\sigma_r/V$ ratio for the gas in the three cases is between 0.1 and
0.2, in agreement with the theoretical expectation,
\equ{sigma}, which is $\sigma_r/V = 0.12-0.17$ for $\delta =0.3$ and $\Qc =
0.67-1.0$. These values are compatible with the observed high-redshift discs
\citep{Genzel06, Forster09} and are significantly
higher than in low-redshift discs.

\subsection{Gravitational Instability}
\label{sec:instability}

\begin{figure}
\vskip 0.1cm
\center
\includegraphics[width =0.36 \textwidth]{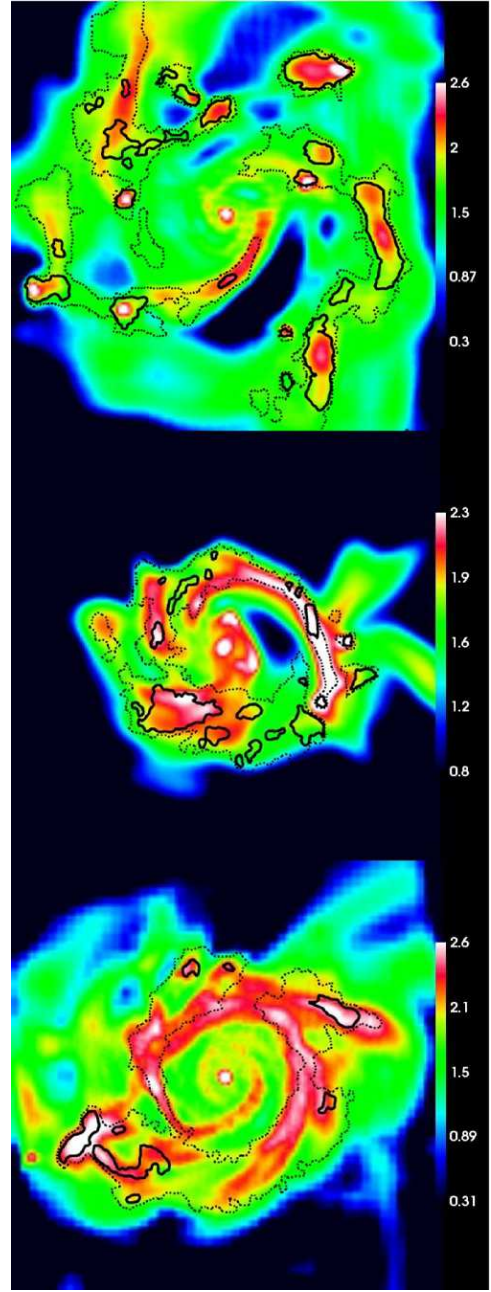}
\caption{
Instability analysis.
Shown are maps of the effective $Q$ parameter in the discs of 
galaxies \tres\ (top), \uno\ (middle) and \dos\ (bottom) overplotted over the 
gas surface density maps of figures \ref{fig:1}, \ref{fig:2} and \ref{fig:3}. 
The value of $Q$ is computed from \equ{Q2} assuming a two-component disc.
The inner, solid contour refers to $Q=1$, and the outer, dashed contour to
$Q=2$.
In all cases, a significant fraction of the disc has $Q <1$, so it is
gravitationally unstable for axi-symmetric modes.
The larger regions where $1<Q<2$ can be subject to higher modes of instability.}
\label{fig:Q}
\end{figure}

\begin{figure*}
\includegraphics[width =0.9\textwidth]{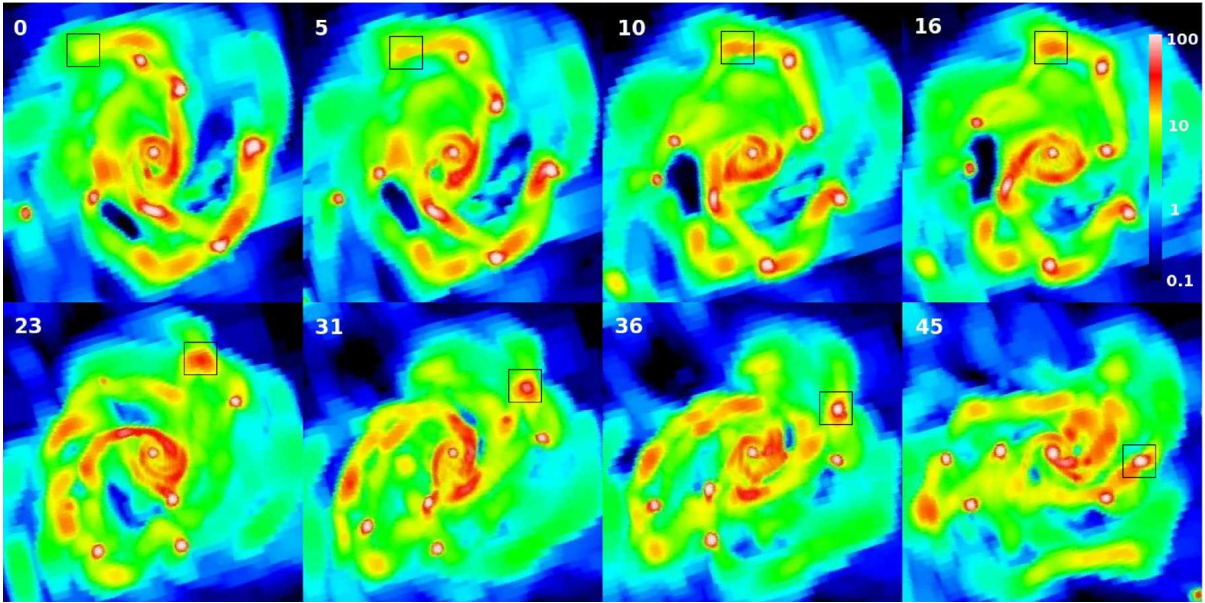}
\caption{Formation and evolution of a clump.
Shown is a time sequence of disc \tres\ near $z=2$.
Each snapshot shows the face-on view of the maximum 3D gas density along
the line of sight in units of Hydrogen atoms cm$^{-3}$.
The size of each panel is $8\kpc$.
A black square of size 1 kpc marks the position of 
one specific clump as it orbits the galaxy.
The clump originates as a small-amplitude perturbation in an unstable
part of the disc and collapses to virialization within a dynamical time.
}
\label{fig:formation}
\end{figure*}

 \begin{figure*}
\includegraphics[width =0.9\textwidth]{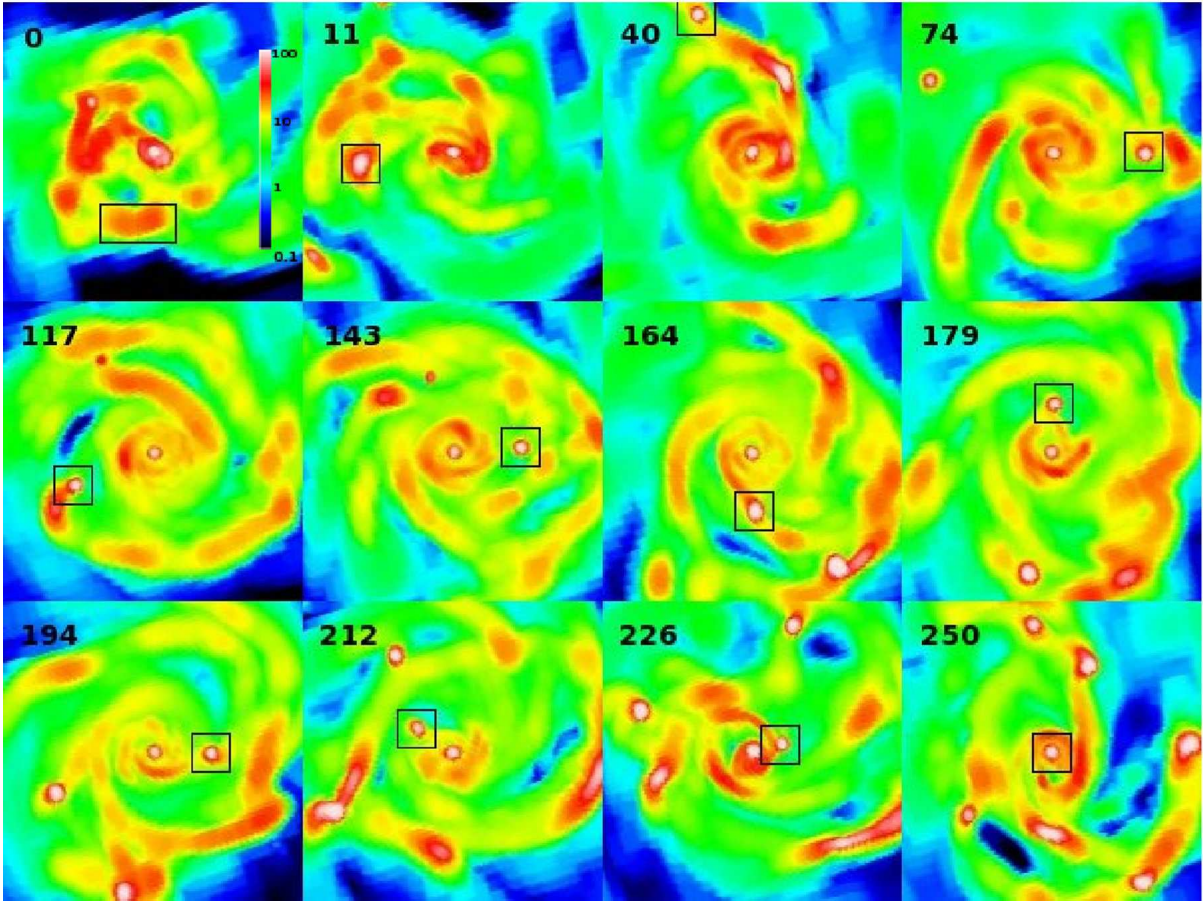}
\caption{Clump Migration into a bulge.
Shown is a time sequence of disc \tres\ near $z =2 $.
Each snapshot shows the face-on view of the maximum 3D gas density along
the line of sight in units of Hydrogen atoms cm$^{-3}$.
The size of each panel is $6\kpc$.
A black square of size 1 kpc marks the position of one specific clump.
The clump forms, virializes, and spirals in. It coalesces with the central
bulge in $250\Myr$, after completing about four rounds.}
\label{fig:migration}
\end{figure*}

Figure~\ref{fig:Q} displays face-on maps of the local effective Q parameter,
computed following \equ{Q2} assuming a two-component disc of gas and stars.
At each point in the disc, we used \equ{Q} separately for each component
to compute $\Qg$ and $\Qs$ from the corresponding values of $\Sigma$ and
$\sigma_r$ and the common value of $\Omega$.
One can see that in all cases, a significant fraction of the disc is unstable
with $Q<1$. These fractions are 33\%, 21\% and 13\% for \tres, \uno\ and \dos\
respectively.
We also note that in galaxies \tres\ and \uno\ the most unstable
regions define extended rings, similar to the rings of star formation observed
in some of the SFGs \citep{Genzel08}.

The interpretation of the $Q$ values obtained here from the simulated
discs is not straightforward because they were measured form the highly
perturbed discs while the linear analysis refers to the slightly perturbed
disc at the onset of instability. 
We see that while the regions where $Q<1$ naturally include all the dense
clumps and transient perturbations, they also extend to regions that seem 
less perturbed, probably on their way to develop nonlinear perturbations. 

Galaxy \tres\ is the ``most unstable" case, reflecting  
its high gas fraction and its low rotational velocity.
Galaxy \dos\ is the most stable case, with a low gas fraction and a high
rotational velocity. 
Still, galaxy \dos\ develops the most pronounced sheared features of the sample.
About 60-70\% of each disc have a Toomre parameter $Q<2$, 
capable of developing high-order instabilities.
It is interesting to note that these discs are highly unstable
despite the dominance of the stellar component. As expected from \equ{Q2},
the stellar component of the disc contributes to the local self-gravity 
that drives the instability, and helps destabilizing the disc.

\begin{figure*}
\center
\includegraphics[width =0.7\textwidth]{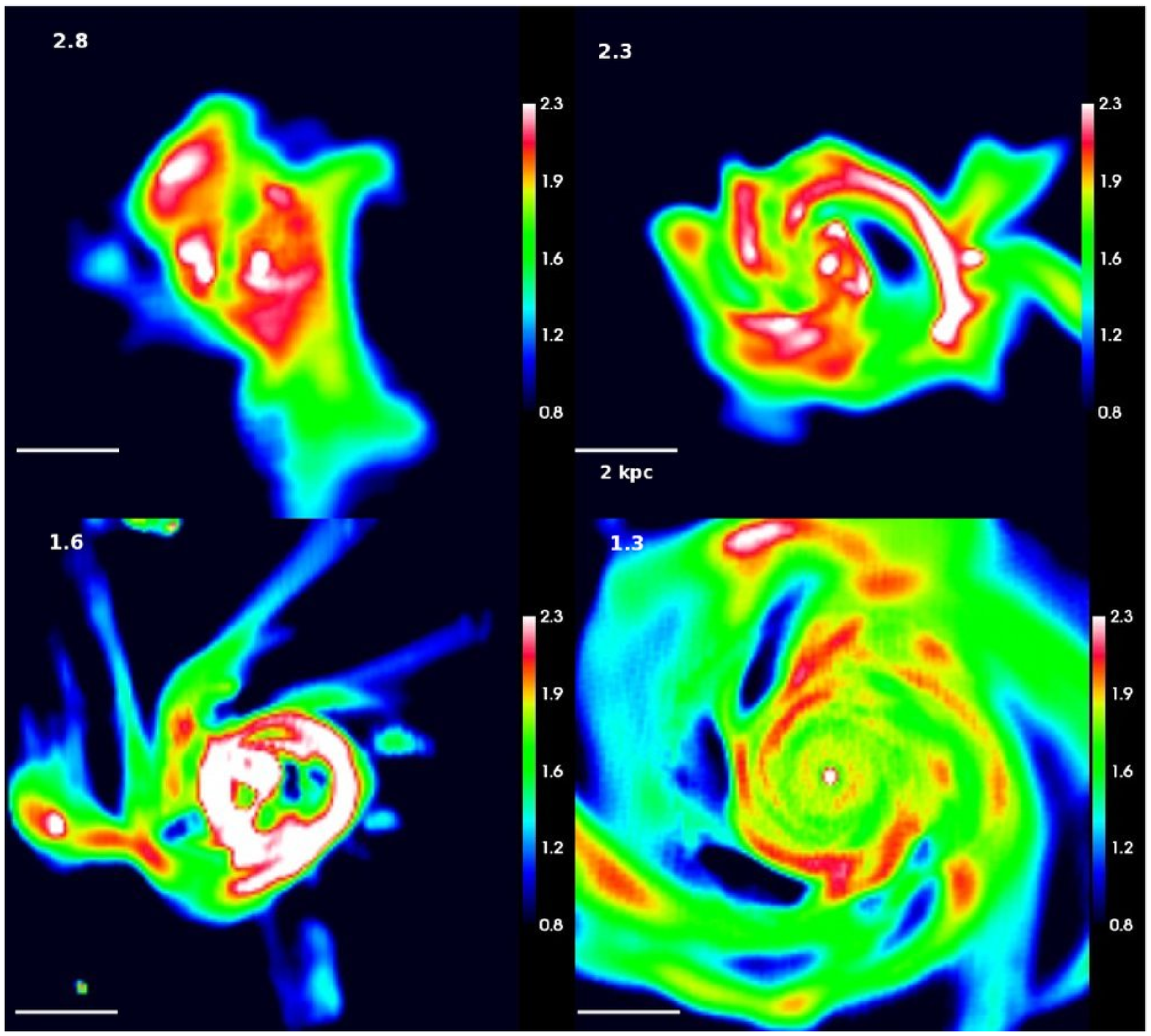}
\caption{
Evolution of galaxy \uno\ in steady state over cosmological times.
Shown are face-on snap-shots of gas surface density at
$z=2.8, 2.3, 1.6, 1.3$ (from left to right) 
in units of log($\msun {\rm pc}^{-2}$). 
The box side is 10 kpc and the color code is the same in the three panels.
These sequence demonstrates that this galaxy maintains a gravitationally 
unstable disc for at least 2-3 Gyr as it is intensely fed by cosmological 
cold streams. This steady-state ends near $z=1.3,$ when the disc is stabilized
by the presence of a dominant bulge --- an example of ``morphological
quenching".}
\label{fig:steady}
\end{figure*}

\subsection{Evolution of Clumps and Spheroid Growth}
\label{sec:evo}

Figure~\ref{fig:formation} follows the formation of an individual clump 
in the disc of galaxy \tres\ near $z\simeq2$. It starts from a small-amplitude 
perturbation in a rather smooth region of the outer disc. 
During the initial 10~Myr, it is tangentially elongated due to shear.
By 30~Myr it has turned into a high-contrast, round, bound and virialized clump.
This is comparable to the dynamical time of this disc, 
$\td = \Rd/V \simeq 31 \Myr$, or the free-fall time for a medium of
$n \simeq 4\cmc$.  

Some of the features seen in these images are transients, where 
the strong shear prevents gravitational collapse. 
Such a transient can be seen in the bottom panels, rotating from
left to top and starting to decay by the last snapshot.
Similar elongated structures in the same galaxy are seen 
in the top panel of figure \ref{fig:1}. They typically grow and decay
on a dynamical timescale. 

Figure~\ref{fig:migration} follows the evolution of an individual gas clump in
galaxy \tres\ for a period of $250\Myr$, 
from a pre-collapse cloud to its final coalescence into the bulge. 
At $t=0$, we see an elongated broad perturbation.
At $t=40\Myr$, after about one dynamical time, the clump seems collapsed and
virialized.  As the clump rotates about the galaxy centre it spirals inward, 
while keeping its size and appearance roughly the same. By $t=250\Myr$,
after circling the centre about four times, the clump has coalesced with the 
central bulge. The migration time from the outer disc to the centre is 
$\sim 8 \td$, very similar to the prediction in \equ{tm}, 
based on 
the gravitational interactions within the disc discussed in \se{theory}.
The SFR inside the clumps starts of at a level of several $\sy$
and it gradually declines as the gas is substantially depleted.
While most of the clump mass turns into stars before the migration is over, 
the clump is still gas rich when it coalesces with the bulge, 
leading to a compact bulge as in a wet merger.

The migration of clumps to the centre contributes to the growth of the 
central spheroid. 10-20\% of the disc mass flows into the bulge in ten
dynamical times, in agreement with the estimated migration time by clump
encounters, \equ{tm}, and the fraction of the disc mass in 
the clumps, $\alpha \simeq 0.1-0.2$.
This is equivalent to one or two mergers of mass ratio 1:10 every 0.3 Gyr,
which is more than expected on average for standard mergers of satellite 
clumps that come with the streams from outside the galaxy
\citep{dekel09}.
Thus, the mass inflow in the disc contributes significantly to the 
early growth of spheroids. The bulge growth starts as soon as the disc 
becomes unstable, and it continues for several Gyr in a near steady state, 
as predicted by DSC.  In this steady state, the continuous gas supply is 
replenishing the draining disc and the comparable bulge and disc keep 
growing in a similar pace.

\subsection{Steady State and Morphological Quenching}
\label{sec:steady}

\Fig{steady} shows face-on snapshots of the gas disc of galaxy \uno\
from $z=2.8$ to $z=1.3$.
During most of this period of $2.6 \Gyr$, the disc appears to be gravitationally
unstable, with transient features and bound clumps, similar to its appearance
at $z=2.3$ as analyzed in detail above. This indicates a near steady state,
as predicted by DSC.
During this period, the gas supplied by the cosmological cold streams
replenishes the gas that is being drained into the bulge through 
clump migration and angular-momentum transport.
The value of $\delta$ at $z=2.8,2.3,1.6$ is $0.23, 0.27 ,0.25$ respectively, 
demonstrating that this steady state is indeed characterized by
a roughly even distribution of mass among the disc, bulge and dark matter halo 
components within the disc radius.
A similar steady state is also seen in the other two galaxies.
We conclude that the three galaxies simulated here represent the typical
stream-fed galaxies observed as rapid star formers at redshifts 3 to 1.5.
A clumpy disc with a smaller bulge could appear within the
first 0.5~Gyr after the onset of instability, perhaps following a period of 
stability due to a dense dark-matter halo, low accretion rate, 
or high turbulence.

However, one can see in \fig{steady} that by $z=1.3$, 
galaxy \uno\ is no longer in the steady-state phase.
The disc appears to be much smoother than before, showing only minor
perturbations of much lower contrast. The disc fraction at this time is 
only $\delta=0.2$, indicating that the disc is being stabilized by the 
presence of a dominant stellar bulge, constituting a significant fraction of
the total stellar mass of $\Ms \simeq 10^{11} \Msun$.  
As a result, the SFR in the disc is 
reduced to $\sim 5 \sy$, about a third of the average cosmological accretion 
rate at this epoch. This is an example of ``morphological quenching" 
of star formation, as discussed and demonstrated in \citet{Martig09}. 
A similar phenomenon occurs in galaxy \tres.
We note that the buildup of the stabilizing bulge in these galaxies 
is mostly due to mass inflow in the disc while in the case of \citet{Martig09}
it is predominantly a result of external mergers. This demonstrates that 
morphological quenching can be obtained in different ways.

We are not in a position to address the properties of the low-redshift
descendents of our three simulated galaxies because the simulations
were stopped at moderate redshifts, where they became too costly.
The last snapshots available for galaxies \tres\ and \uno\ are at
$z=1.3$, and for galaxy \dos\ at $z=1.9$.
The appearance of a dominant stellar spheroid and a minor stable
disc of stars and gas in galaxies \uno\ and \tres\ hints that these 
galaxies may end up as early-type galaxies with low SFR. 
On the other hand, as demonstrated by \citet{Martig09}, such galaxies
may eventually re-grow a sufficiently massive gas disc that becomes unstable 
again, and thus turn back into a blue, star-forming, late-type galaxy.

\section{Artificial Fragmentation}
\label{sec:artif}

\begin{figure}
\center
\includegraphics[width =0.45 \textwidth]{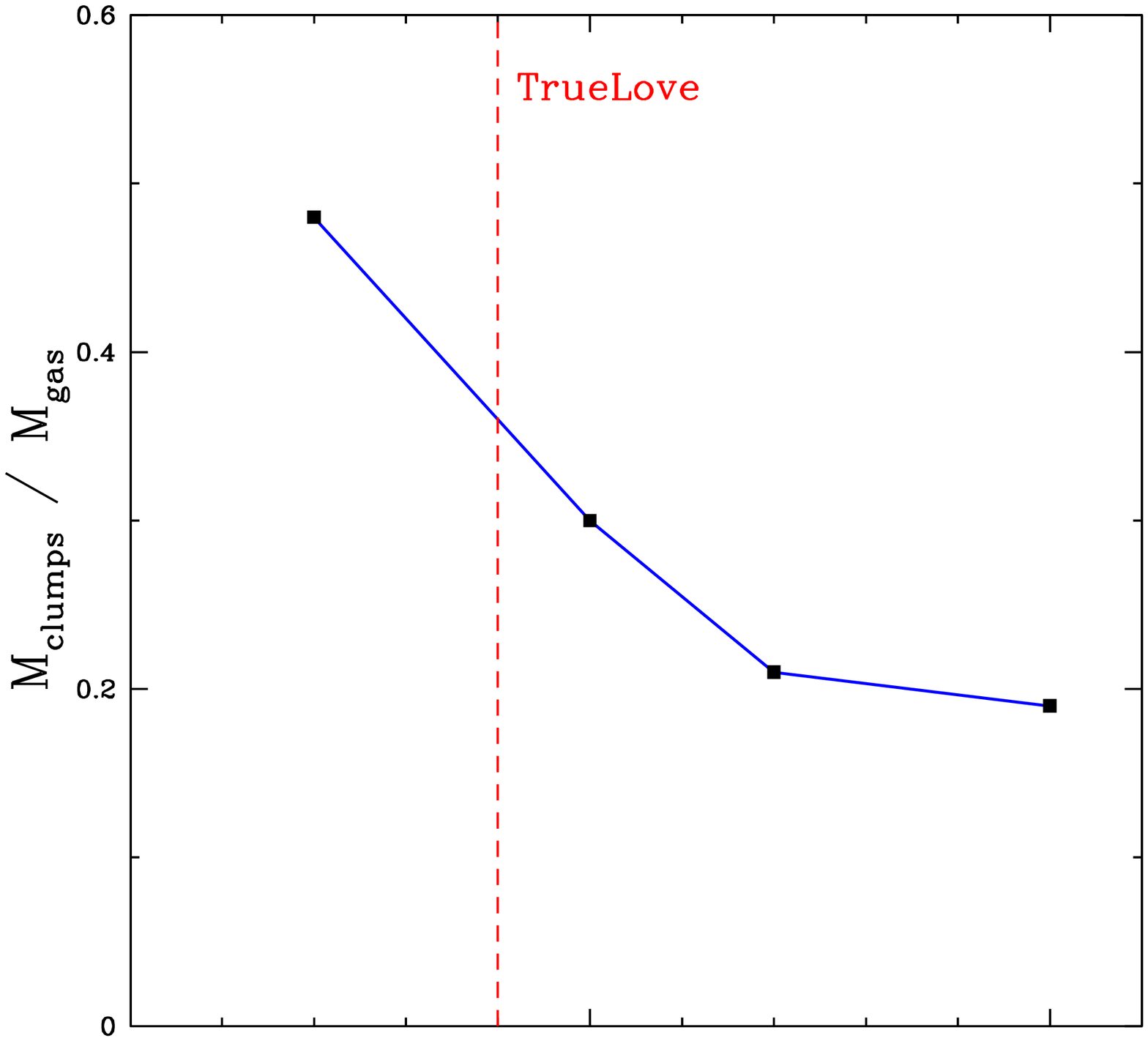}
\includegraphics[width =0.45 \textwidth]{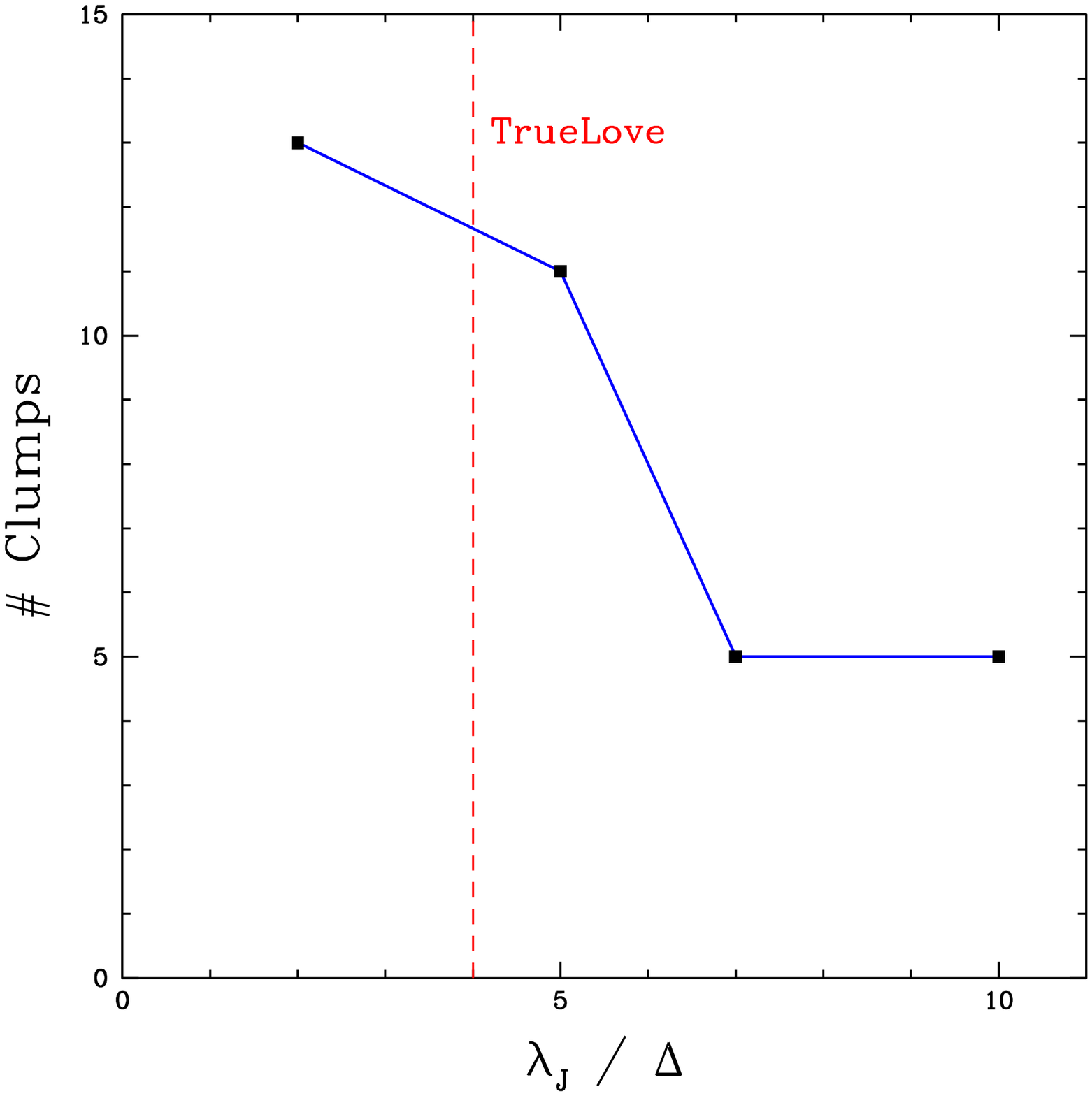}
\caption{
Testing artificial fragmentation.
Shown is the mass fraction in clumps (top panel), and the number of 
clumps (bottom panel) for simulations with different values of $\Nc$.
Convergence is indicated at $\Nc \geq 7$, and artificial fragmentation
becomes gradually more severe as $\Nc$ becomes smaller below $\Nc = 7$.
At the Truelove value, $\Nc=4$, the mass in clumps is larger than the
base value at $\Nc \geq7$ by a factor of $\sim 1.8$ and the number of 
clumps is higher by a factor of $\sim 2.3$. The figure indicates that 
$\Nc\gsim 7$ is the recommended value for a pressure floor that would 
safely eliminate artificial fragmentation.}
\label{fig:art}
\end{figure}


If the Jeans length, $\lamj$, is comparable to the resolution
scale of a grid-based code, i.e., the smallest cell size $\Delta$,
one expects artificial fragmentation on that scale.
This is because the code fails to properly follow the pressure gradients
that are supposed to support a clump on that scale against gravitational
collapse.
\citet{Truelove} have shown that 
the artificial fragmentation could be prevented 
if the Jeans length is always resolved by more than 4 resolution
elements.\footnote{A
similar criterion is necessary in SPH simulations, where
the Jeans length has to be resolved by more than a given number of
SPH kernels \citep{Bate97}.}
If we define a Jeans number by
\be
N_{\rm J} \equiv \frac{\lamj}{\Delta} \, ,
\ee
it has to be kept above a critical value, $\Nc=4$. 
The Truelove value was obtained based on simulations that lasted only
a few free-fall times, 
so it is possible that a more stringent criterion may
be required when the system is followed for longer periods, typical of 
cosmological simulations.

Our current cosmological simulations do not resolve the thermal Jeans length 
of the coolest gas, $T\sim 300$K, so one must act to prevent artificial 
fragmentation.  For this purpose we have implemented an artificial pressure 
floor via a procedure that is becoming standard
\citep{Machacek01, RobertsonKravtsov, Agertz09b}.
For gas with adiabatic index $\gamma$, the speed of sound is related to the
pressure and density via $\cs^2=\gamma P/\rho$, so the Jeans length is given by
\be
\lamj^2 = \frac{\pi \gamma P}{G \rho^2} \, .
\ee
In order to set the minimum $N_{\rm J}$ to a 
desired $\Nc$, 
we thus set the pressure floor at
\be
\Pf = \frac{G \rho^2 \Nc^2  \Delta^2}{\pi\gamma } \, .
\ee
The pressure in the Euler equation is replaced by an effective pressure
that is 
set to $\Pf$ when it would have otherwise obtained a lower value.
We adopt $\gamma=5/3$.
 

In order to verify the value of $\Nc$ necessary for preventing artificial
fragmentation in our cosmological discs, we re-ran the simulation of case 
\uno\ for the last 400~Myr prior to $z=2.3$ with different values 
of the pressure floor, or $\Nc$. 
The duration for this test is about 22 dynamical times, which should
be sufficient for studying the fragmentation that develops on one or a few
dynamical timescales. 
The corresponding values of $\Nc$ were chosen to be 2, 5, 7 and
10, namely from half the Truelove value of $\Nc=4$ to more than twice
above it.
There is no point testing higher values of $\Nc$ because the 
corresponding Jeans length would have become comparable to the disk thickness 
and thus prevent the real disk fragmentation that we are after.

Figures \ref{fig:art} shows the fraction of gas in clumps and the number of
clumps as a function of $\Nc$.  The latter is measured as the number of
density peaks with $n > 25 \cmc$, and the trend is found to be insensitive
to the exact threshold chosen.
Below $\Nc = 7$, we see a strong increase in both the mass in clumps and the
clump number as a function of decreasing $\Nc$.
On the other hand, for $\Nc$ in the range $7-10$, we see a convergence of 
the two clump indicators into constant values.
At the Truelove value, $\Nc=4$, both indicators have values
about twice as large as their base values at $\Nc=7-10$.
The similar behavior of the two indicators implies that the artificial
fragmentation at small values of $\Nc$ is expressed in terms of the 
appearance of new clumps more than in the growth in mass of the real clumps
that exist at large values of $\Nc$. 

We conclude that in order to avoid artificial fragmentation, the Jeans length 
must be resolved by 7 elements or more. 
We adopt $\Nc=7$ as our default case.
With such a pressure floor, the clumps are likely to be real and their global 
properties are expected to be reliable. They are supported by turbulent motions
in the disc and their size is set by the turbulence Jeans length.
On the other hand, the detailed internal properties of the clumps, including 
density profile and substructure, are subject to the limited resolution.
This internal structure can be resolved only in simulations of higher
resolution.

\section{Discussion and Summary}
\label{sec:disc}


We analyzed high-resolution AMR simulations of three galaxies in haloes 
of total mass $\sim 5\times 10^{11} \msun$ and baryonic mass 
$\sim 5\times 10^{10} \msun$  
at $z \simeq 2.3$, selected quite randomly from a cosmological simulation.
Each of these cases show a clumpy, thick and turbulent disc of gas and stars 
extending to $(3-6)\kpc$ in radius and accommodating a central stellar 
bulge of comparable mass within the inner $(1-2)\kpc$.
The discs show a very irregular morphology, with elongated transient features
and a few bound massive clumps. 
The simulations reveal that these are naked 
clumps, not associated with dark matter haloes, and formed in-situ in the disc
by gravitational instability. 
The simulated clumpy morphology is remarkably similar to the typical
morphology of high-redshift SFGs as observed from different angles of view.

The simulations indeed confirm many of the detailed predictions 
outlined in \citet{dsc} based on a straightforward Toomre instability 
analysis in a cosmological context. They demonstrate that 
a large fraction of the massive galaxies at $z \simeq 2-3$ are made of a
gravitationally unstable disc with a stellar bulge of a comparable mass,
corresponding to a disc-to-total mass ratio of $\delta \simeq 0.3$ in 
\equ{delta}.  
A significant fraction of the disc has an effective value of $Q\leq 1$,
typically defining an extended ring of maximum instability.
The in-situ giant clumps are $\sim 10^8\msun$ each, namely
on the order of one percent of the disc mass, 
in agreement with \equ{Mc}. They involve 10-20\% of the disc mass,
as assumed in DSC. 
The clumps migrate to the disc centre in about $250\Myr$, on the order
of ten disc dynamical times, where they coalesce into a bulge.
The mass inflow in the disc to the bulge contributes significantly to the 
spheroid growth, adding to the effect of mergers of external galaxies.

In our simulations, the clumps survive bound until they merge into the bulge.
It has been argued that in reality the clumps are disrupted
on a dynamical timescale by radiative feedback from newly formed stars
\citep{Murray09}. 
In an associated paper (Krumholz \& Dekel, in preparation),
it is shown that the ejected fraction of gas from the giant clumps in 
massive disks at high redshift is limited
to less than $10\%$ provided that the SFR efficiency is $\sim 0.01$, i.e. it
obeys the Schmidt-Kennicutt relation between SFR and gas density 
\citep{Kennicutt98}, which is also favored theoretically 
\citep{Krumholz05, Krumholz07}. The clumps could be disrupted only if the SFR efficiency was larger than its
local value by an order of magnitude.
 
The SFR at $z=2.3$ is at the level of $\sim 40\sy$, which is in the ball 
park of the expected cosmological gas accretion rate along cold streams 
into haloes of comparable masses \citep{dekel09}. This corresponds to 
more than $100 \sy$ in the massive SFGs that reside in haloes more massive
than $10^{12}\msun$. We recall, though, that the simulated SFR at $z \sim 2.3$ 
may be slightly underestimated due to the limitations of 
the sub-grid recipes for star formation and feedback.
Given these limitations,
about half the star formation seems to occur in the disc clumps. 
The star formation within each clump extends to $\sim 200\Myr$, 
or several disc dynamical times, which is comparable to the timescale 
for clump migration.
If the star-formation efficiency is $\sim 1-2\%$ 
as in the molecular clouds of today's galaxies \citep{Krumholz09},  
then it must occur in denser sub-clumps (DSC), 
which are not resolved in our current simulations.

Our simulations confirm that the disc is continuously supplied by fresh
material through cold streams that penetrate the hot halo along 
the dark-matter filaments of the cosmic web.
In a companion paper, we use the simulations to predict the observational 
appearance of these cold streams as extended Lyman-alpha blobs 
\citep{Goerdt09}.
Despite the very high resolution, 
the streams in our three galaxies are rather smooth; 
the fraction of the incoming stream mass in clumps with densities higher 
than $0.1\cmc$ ranges from 7\% to 25\%, in comparison with the average
value of 30\% estimated by \citet{dekel09} for massive galaxies at 
similar redshifts. As argued by DSC, this smoothness of the incoming mass
is a key for maintaining the disc sufficiently dense and cold 
without growing a stabilizing bulge, such that the disc is kept 
gravitationally unstable.
The continuous supply of gas keeps this configuration of an unstable disc
and a comparable bulge in a near steady state that lasts for 2-3 Gyr.

At least two of the galaxies simulated here become rather
stable before $z=1$; 
they become dominated by a stellar spheroid that stabilizes the remaining disc,
prevents its fragmentation into giant clumps, and suppresses star formation
to quiescent levels. This is a demonstration of the morphological-quenching
mechanism discussed by \citet{Martig09} and seen in the simulation
of \citet{Agertz09b}.  

Two of the simulated galaxies show a mass fraction of 
galactic baryons inside the dark-matter halo which is $\sim 30\%$ 
of the cosmological baryonic fraction, while the third galaxy has a baryon
fraction comparable to the universal value. The comparison with 
the low baryonic fractions indicated from observations at low redshift 
(e.g., F.~Prada, private communication)
is beyond the scope of this paper, because our simulations are currently
limited to high redshift. We anticipate that low baryonic fractions can 
be obtained if the baryon accretion is suppressed at low redshifts
while the dark-matter haloes keep growing.

While the current simulations are state of the art, they are still limited
in many ways. One limiting issue is the relatively low gas fractions of 
$8-35\%$ in the simulated discs compared to the values of $30-60\%$
indicated for the observed galaxies. Although the observational estimates are 
indirect and therefore uncertain, it is possible that our current simulations 
slightly overestimate the star formation rate at earlier times, 
possibly as a 
result of underestimating the suppression of SFR by stellar and supernova 
feedback. This may lead to an underestimate of the gas fraction and SFR at 
$z = 2.3$. Fortunately, this possible limitation does not prevent the robust 
appearance of strong instability in the high-redshift discs at $z \geq 1.5$,
so the simulations do produce galaxies that mimic the main features
of the observed SFGs and the theoretical predictions.

Another caveat is associated with the mandatory incorporation of a pressure
floor. While it serves the purpose of preventing artificial fragmentation,
it necessarily smoothes the thermal-pressure gradients in high density regions. 
Only simulations with a resolution better than $10 \pc$ could help us
directly evaluate the side effects that may be introduced by this 
pressure floor on small scales. 
However, the tests presented in \se{artif} demonstrate that the global
properties of the clumps are stable against variations of the pressure floor 
in the regime where artificial fragmentation is suppressed.
This is because the gravitational collapse of the clumps is balanced
by velocity dispersion rather than by thermal pressure gradients.
The velocity associated with the artificial pressure is $\sim 8 \kms$
compared to the turbulent velocity of $20-30 \kms$.
In practice, over most of the simulated unstable disks, the Jeans length 
associated with the overall pressure is significantly larger than the 
artificial minimum imposed at 7 cell sizes.
 
Our main findings from the clumpy galaxies at high redshift are consistent with 
the parallel findings of \citet{Agertz09b}, who used a simulation with 
a similar resolution but a different AMR code.
They also see clump formation in massive spiral arms driven by disc
instability.  However, they find about twice as many bound clumps, ranging
to smaller masses
and preferentially populating the outer disc, which seem not to resemble
the observed SFGs as well as our simulations.
Furthermore, they detect significant small-scale
fragmentation in the initially smooth, cold streams, where we see no such
effect. This is probably a result of some degree of artificial 
fragmentation in \citet{Agertz09b} due to the lower pressure floor adopted
in their simulation, (\se{artif}). 
When resolving the Jeans length with only 4 resolution elements,
following the Truelove criterion, we find that artificial fragmentation tends 
to occur in regions where the gas velocity dispersion is low such that the 
Jeans length is strongly affected by thermal pressure, as in the coherent 
cold streams. 
On the other hand, when we resolve the Jeans length by 7 resolution elements, 
we prevent such artificial fragmentation in our simulations.
Still, the many qualitative similarities between the findings of
\citet{Agertz09b} and ours indicate that we both capture the real
key features of high-redshift SFGs.  
  
We conclude that zoom-in simulations of galaxies drawn quite randomly
from a cosmological volume reproduce clumpy discs and bulges similar to 
the observed star-forming galaxies at high redshift. The simulations 
demonstrate the validity of the simple theoretical picture where the 
continuous gas supply by cold streams drives a self-regulated gravitational 
instability in a cosmological steady state that tends to last 
for several Gyr.

\section*{Acknowledgments}

The computer simulations were performed at the National Energy Research 
Scientific Computing Center (NERSC), Lawrence Berkeley National Laboratory,
PI J. Primack.
We acknowledge stimulating discussions with Andi Burkert,
Bruce Elmegreen, Reinhard Genzel, Tobias Goerdt, Anatoly Klypin, Mark Krumholz,
Doug Lin, Joel Primack, Re'em Sari, Kristen Shapiro,
Amiel Sternberg and Linda Tacconi.
This research has been partly supported by an ISF grant,
by GIF I-895-207.7/2005, by a DIP grant, by a France-Israel Teamwork
in Sciences, by the Einstein Center at HU,
and by NASA ATP NAG5-8218 at UCSC.
DC is a Golda-Meir Fellow at HU.

\bibliographystyle{mn2e}
\bibliography{ceverinoC3}

\bsp

\label{lastpage}

\end{document}